\documentstyle[eqsecnum,aps,preprint,epsfig]{revtex}
\preprint{NCL93-TP18,gr-qc/9412075}
\tighten
\begin{document}

\title{ ONE-LOOP QUANTUM GRAVITY \\
 IN SCHWARZSCHILD SPACE-TIME}

\author{Bruce P.~Jensen}
 \address{ Facult\'e des Sciences,
Universit\'e de Corse, 20250 Corte, France \\
 {\rm and} Dept.~of
Physics, University of Newcastle, Newcastle upon Tyne NE1 7RU,
UK\ $^*$\\
email: B.P.Jensen@newcastle.ac.uk}

\author{John G.~Mc Laughlin}
 \address{ Dept.~of
Physics, University of Newcastle, Newcastle upon Tyne NE1 7RU,
UK\ $^\dagger$}

\author{Adrian C.~Ottewill}
\address{ Mathematical Institute,
University of Oxford, 24-29 St.~Giles', Oxford OX1 3LB, UK\\
 {\rm and} Dept. of Mathematical Physics,
University College Dublin,  Dublin 4, Ireland.\\
email: ottewill@relativity.ucd.ie}

\date{\today}
\maketitle

\widetext
\begin{abstract}
The quantum theory of linearized perturbations of the gravitational
field of a
Schwarzschild black hole is presented. The fundamental operators  are
seen to
be the perturbed Weyl scalars $\dot\Psi_0$ and $\dot\Psi_4$ associated
with
the Newman-Penrose description of the classical theory. Formulae are
obtained
for the expectation values of the modulus squared of these operators in
the
Boulware, Unruh and Hartle-Hawking quantum states. Differences between
the
renormalized expectation values of both
$\bigl| \dot\Psi_0 \bigr|^2$ and $\bigl| \dot\Psi_4 \bigr|^2$
in the three quantum states are evaluated numerically.

\end{abstract}

\narrowtext
\section{INTRODUCTION}
\label{intro}
In this paper we shall study quantized, linear perturbations of the
gravitational field of a Schwarzschild black hole. Since this
represents a study of the one-loop approximation to a theory of quantum
gravity, it should  provide useful insights into what the full theory
should look like. In addition, there is  a growing body of evidence
which suggests that the influence of non-conformally invariant quantum
fields, foremost of which is linearized gravity, strongly dominates
that of conformally invariant quantum fields in the neighborhood of a
space-time singularity. For example it has been shown
recently\cite{AMO} that the value of the renormalized energy density
for the graviton field in the vicinity of a conical singularity in an
otherwise flat space-time (the idealized cosmic string) is roughly ten
times that for the electromagnetic field and one hundred times that for
the conformally coupled scalar field. A knowledge of how  gravitons
behave near the singularity at the center of a black hole is therefore
also likely to be crucial to our understanding of quantum gravity.

The classical theory of linearized perturbations of a Schwarzschild
black hole was developped by Regge and Wheeler,\cite{RW}
Zerilli\cite{Z} and others.  Here we follow the approach of
Teukolsky,\cite{T1} who gave a complete set of solutions to the
perturbation equations for a Kerr black hole  within the Newman-Penrose
formalism. In this formalism  the system is  described by  two
quantities --- the perturbed Weyl scalars $\dot\Psi_0$ and $\dot\Psi_4$
(here and in the following, a dot indicates a linearly perturbed
quantity). These are the field variables of interest describing the
semi-classical theory, since

\noindent{(a)} they carry in their real and imaginary parts
information on all the dynamical degrees of freedom of the perturbed
field;

 \noindent{(b)} they are invariant under both gauge transformations and
infinitesimal tetrad transformations which leave the metric
perturbation
intact,\cite{T2} as befits a physical quantity;

 \noindent{(c)} they are simply expressed (albeit in a
particular gauge) in terms of the metric perturbation which
provides the most direct route to quantization;\cite{C} and

 \noindent{(d)} they directly measure the energy flux of
classical perturbations of the black hole across the horizon and at
infinity.\cite{T2,HH}

In this work our main concern will be to extend the work of Candelas
{\it et al.},\cite{CCH} by first presenting formulae for the
expectation values of
$\bigl| \dot\Psi_0 \bigr|^2$ and $\bigl| \dot\Psi_4 \bigr|^2$ with
respect to the three
physically relevant quantum states for the Schwarzschild black hole,
namely the   Boulware, Unruh and  Hartle-Hawking states, and then
evaluating these expressions numerically.
Since $\bigl| \dot\Psi_0 \bigr|^2$ and $\bigl| \dot\Psi_4 \bigr|^2$ are
quadratic in the field variables their expectation values will be
infinite; however differences between the expectation values of either
$\bigl| \dot\Psi_0 \bigr|^2$ or $\bigl| \dot\Psi_4 \bigr|^2$ with
respect to the various quantum states mentioned
above are finite and these differences are the objects which we will
calculate. (Indeed, some would argue that such differences are more
likely to be of physical significance than the results obtained from
renormalization procedures.) In the process of achieving these main
goals we also fill several gaps in the literature concerning quantized
perturbations of black holes; in particular we present both
low-frequency analytic formulae  and a set of Wronskian relations for
the reflection and transmission coefficients associated with the
theory.

The format of the paper is as follows: the two sections immediately
following this introduction summarize the essentials of what is already
known about the classical and semi-classical theories of linear
perturbations of black holes.  Thus in Sec.~\ref{classical} we present
Teukolsky's complete set of classical solutions for the perturbed Weyl
scalars. Since Schwarzschild space-time is static and spherically
symmetric, the temporal and angular dependences of each of these
solutions are given by  $\exp(-i\omega t)$ and spherical harmonic
functions respectively; the radial component however is a solution of a
complex, second order, ordinary differential equation  for which no
closed-form general solution exists. We discuss the behavior of the
radial component  both  as one approaches the event horizon and
spatial infinity. We end Sec.~\ref{classical} by writing down formulae
for the classical energy flux due to the perturbations across the event
horizon and at infinity.

Sec.~\ref{semi-classical} is a review of the semi-classical theory.
We present the complete set of solutions for the metric perturbations
due to Chrzanowski,\cite{C} and verify that these yield Teukolsky's
complete set of perturbed Weyl scalars. The question of orthonormality
of the mode set is rigorously addressed, the perturbations are
quantized  and expressions for differences between the expectation
values of $\bigl| \dot\Psi_0 \bigr|^2$ and
$\bigl| \dot\Psi_4 \bigr|^2$ in the  Boulware, Unruh and
Hartle-Hawking  states are presented in terms of the radial functions.

In Sec.~\ref{powerseries}  we prepare the expressions obtained for
the various expectation values for numerical evaluation. The first step
is to derive a power series representation for the general solution to
the radial equation; the particular radial solutions occuring in the
expectation values may then be substituted for these power series,
weighted by appropriate constants called reflection and transmission
amplitudes by virtue of the analogy with classical scattering.

Before the differences between the expectation values can be evaluated
numerically, the reflection and transmission amplitudes must be
determined explicitly. This is achieved in Sec.~\ref{powerseries} by
comparing the power series representations for the particular radial
solutions at large values of the radial coordinate  with their known
asymptotic forms. Graphs are presented for these amplitudes against
frequency and angular number, and the results are checked for both
internal and external consistency; the former by means of Wronskian
relations derived between the coefficients in Appendix \ref{relations}
and analytic
formulae for the amplitudes valid at low frequencies derived in
Appendix \ref{smallomegaapp}; the latter by reproducing Page's\cite{P}
results for the luminosity due to graviton emission from the black
hole.

In Sec.~\ref{asymptotics} we give  an analysis of the
asymptotics of   the expressions for the differences between the
expectation values of the perturbed Weyl scalars.

 In Sec.~\ref{numerical} we present graphs for these differences,
and discuss the
main difficulties encountered in the numerical computation.

Throughout we use geometrized Planck units ($G=c=\hbar=1$)
and follow the sign
conventions of Misner, Thorne and Wheeler.\cite{MTW}

\section{REVIEW OF THE CLASSICAL THEORY}
\label{classical}

For space-times possessing a high degree of symmetry, it is often
possible to
reduce considerably the number and complexity of the equations of
general
relativity by projecting them onto a null, complex tetrad which encodes
that
symmetry. This is the motivation which lies behind the Newman-Penrose
(NP)
formulation of general relativity.\cite{NP}

In the case of a Schwarzschild black hole of mass $M$, whose space-time
is
described by the metric
\[
 g_{\mu\nu}=\mathop{\rm diag} \left[ -{\triangle\over r^2},
{r^2\over\triangle}, r^2, r^2\sin^2\theta
\right]_{\mu\nu} , \qquad \triangle=r(r-2M)
\]
 in terms of the usual spherical polar co-ordinates
 $(t,r,\theta,\phi)$, an
appropriate choice of tetrad  which mirrors the temporal and spherical
symmetry of the space-time is the {\it Kinnersley tetrad}
\begin{eqnarray}
  e_{(1)}{}^\mu &=& \left( {r^2 \over \triangle}, 1,0,0 \right)
  \nonumber \\
e_{(2)}{}^\mu &=& {1\over2} \left( 1, -{\triangle\over r^2},0,0
\right) \label{tetrad} \\
e_{(3)}{}^\mu = e_{(4)}{}^{\mu*} &=&
{1\over \sqrt2 r} \left(0,0,1,{i\over\sin\theta} \right)  .\nonumber
\end{eqnarray}
The simplification afforded by using this tetrad is exemplified by the
fact
that all spin coefficients vanish apart from
\begin{equation}
\rho = -{1\over r} \qquad \mu   =-{\triangle\over2r^3} \qquad \gamma
= {M\over2r^2}
\qquad -\alpha = \beta  = {\cot\theta\over2\sqrt{2}r}    \nonumber
\end{equation}
and all tetrad components of the Weyl tensor are zero apart from
\begin{equation}
 \Psi_2={M\over r^3}.      \nonumber
\end{equation}
(More than one set of conventions exists for the NP formalism; those
employed in this paper are spelled out in Appendix \ref{notation} to
avoid any
confusion.)

Teukolsky\cite{T2} took full advantage of the simplifications afforded
by the NP formalism when investigating perturbations of the
gravitational field of a Kerr black hole. He obtained linearly
perturbed versions  of the NP analogues of the basic equations of
general relativity for the black hole metric and, by working in the
Kinnersley tetrad, was able to decouple the equations for the ingoing
and outgoing radiative parts of the perturbed Weyl tensor, $\dot\Psi_0$
and
$\dot\Psi_4$.  These were found to satisfy the ``master perturbation
equation,'' which, in the Schwarzschild limit, reads
\begin{eqnarray}
  &&\left[ {r^4\over\triangle} {\partial^2\over\partial t^2} + {2rs(r-
3M)\over\triangle} {\partial\over\partial t} - \triangle
{\partial^2\over\partial r^2} - 2(s+1)(r-M) {\partial\over\partial r}
\right.           \nonumber      \\ && \left. -
{\partial^2\over\partial\theta^2} -
\cot\theta {\partial\over\partial\theta} - {1\over\sin^2\theta}
{\partial^2\over\partial\phi^2} - {2is\cos\theta\over\sin^2\theta}
{\partial\over\partial\phi} + s^2 \cot^2 \theta - s \right] \Phi_s = 0
  \label{master}
\end{eqnarray}
where $s=\pm2, \Phi_2\equiv\dot\Psi_0$ and $\Phi_{-2}\equiv
r^4\dot\Psi_4$. It
is remarkable that this equation can also be shown to describe the
behavior of
a test scalar field ($s=0,\Phi_0\equiv\varphi$) or electromagnetic
field
($s=\pm1,\Phi_1\equiv\phi_0,\Phi_{-1}\equiv r^2\phi_2$, where $\phi_0$
and
$\phi_2$ are the tetrad components $F_{\mu\nu} e_{(1)}{}^\mu
e_{(3)}{}^\nu$
and $F_{\mu\nu} e_{(4)}{}^\mu e_{(2)}{}^\nu$ respectively of the
Maxwell
tensor $F_{\mu\nu}$) on the Schwarzschild background. For this reason
we shall
keep $s$ arbitrary whenever possible in the ensuing discussion.

By separating variables in (\ref{master}) the following
complete set of solutions is
obtained:
\begin{equation}
 \Phi_s={\rm e}^{-i\omega  t}\,{}_{s} R_{l\omega}  (r)\,{}_{s}
 Y_l^m(\theta,\phi) \label{phis}
\end{equation}
 where $\omega \in [0,\infty)$ and $l$, $m$ are integers satisfying the
inequalities $l\ge|s|$, $-l\le m\le l$. $\,{}_{s} Y_l^m(\theta,\phi)$
is a
{\it spin-weighted spherical harmonic}, whose relevant
properties are reviewed in
 Appendix B. $\,{}_{s} R_{l\omega}  (r)$
satisfies the ordinary differential equation
\begin{equation}
 \left[ \triangle^{-s} {{\rm d}\over {\rm d}{r}}  \left(
 \triangle^{s+1}
 {{\rm d}\over {\rm d}{r}}
 \right) +{\omega^2 r^4 +2is\omega
r^2(r-3M)\over\triangle}  -(l-s)(l+s+1) \right] \,{}_{s} R_{l\omega}
 (r) = 0. \label{req}
\end{equation}
 The solutions of  (\ref{req}) cannot be obtained analytically in
 closed
form (we shall solve it
 numerically later). However much can be
 learned about their
properties by exploiting the analogy of (\ref{req}) to a classical
scattering
problem, which becomes apparent when we write (\ref{req}) in the
form\cite{T2}
\begin{equation}
 \left[{{\rm d}^2\over {\rm d}{r_*}^2}+\,{}_{s} V_{l\omega}
 \right]\,{}_{s} Q_{l\omega}  =0. \label{scateq}
\end{equation}
Here $r_*$ is the {\it tortoise co-ordinate}
\[
 r_*=r+2M\ln\left({r\over2M}-1\right),
\]
$\,{}_{s} Q_{l\omega}  (r) = \triangle^{s\over2} r \, \,{}_{s}
 R_{l\omega}(r)$ and
 $\,{}_{s} V_{l\omega}  (r)$ is the (complex)
potential
\begin{equation}
\,{}_{s} V_{l\omega}(r)
 =\omega^2 +{2is\omega(r-3M)\over r^2} -{s^2 M^2\over r^4}
-{r-2M \over r^3}\left[
l(l+1) + {2M \over r}\right]. \label{pot}
\end{equation}
 As $r\to\infty$ the potential  (\ref{pot}) can be approximated by
\[
 \,{}_{s} V_{l\omega}(r) \sim \omega^2 +{2i\omega  s\over r}
\]
 so that (\ref{scateq}) possesses solutions which behave like
\[
 \,{}_{s} Q_{l\omega}(r) \sim r^{\pm s} {\rm e}^{\mp i\omega  r_*}
\]
 which gives
\begin{equation}
 \,{}_{s} R_{l\omega}  (r) \sim r^{-1} {\rm e}^{-i\omega  r_*} \qquad
 \hbox{or}
 \qquad r^{-2s-1} {\rm e}^{+i \omega  r_*}  .
\label{simpleasymp}
\end{equation} Similarly, in the limit as $r\to 2M$,
\begin{equation}
 \,{}_{s} V_{l\omega}   \sim \left( \omega-{is\over4M} \right)^2
 \nonumber
\end{equation}
 so that
 \begin{equation}
 \,{}_{s} Q_{l\omega}  (r) \sim \triangle^{\pm{s\over2}}
 {\rm e}^{\pm i\omega  r_*} \nonumber
\end{equation}
 and
\begin{equation}
\,{}_{s} R_{l\omega}  (r) \sim \triangle^{-s} {\rm e}^{-i\omega  r_*}
 \qquad \hbox{or} \qquad
 {\rm e}^{+i \omega  r_*} \quad . \label{asymp2M}
\end{equation}
To end this section, we justify the assertion made in the Introduction
that
the perturbed Weyl scalars $\dot \Psi_0$ and $\dot \Psi_4$ measure the
energy
flux of classical perturbations of the Schwarzschild black hole across
the
horizon and at infinity.

Hartle and Hawking\cite{HH} have shown that the energy flux transmitted
across
the horizon of a fluctuating classical black hole is given by the
formula
\begin{equation}
{{\rm d}^2 E^{\rm hor}_\omega  \over {\rm d}t {\rm d}\Omega }
 ={ M^2 \over \pi }
\bigl|  \dot\sigma_{\rm H} (2M) \bigr|^2 \nonumber
\end{equation}
 where $\dot\sigma_{\rm H}(r)$ is the perturbed shear of the null
 congruence
$\left( e_{(1)}{}^\mu \right)_{\rm H}$ which generates the future
horizon,
the calculation having been performed in the {\it Hartle-Hawking
tetrad} which
is defined as follows in terms of the Kinnersley tetrad
(\ref{tetrad}):
\[
  \left( e_{(1)}{}^\mu \right)_{\rm H} = {\triangle\over 2 r^2}
e_{(1)}{}^\mu  , \qquad\left (e_{(2)}{}^\mu \right)_{\rm H}
= {2r^2\over\triangle}
e_{(2)}{}^\mu , \qquad \left( e_{(3)}{}^\mu \right)_{\rm H} =\left(
e_{(4)}{}^{\mu*} \right)_{\rm H} =e_{(3)}{}^\mu.
\]
 It can also be shown from the perturbed Newman-Penrose equations that
for perturbations of frequency $\omega$  for which $\left( \dot
e_{(1)}{}^\mu
\right)_{\rm H} \propto \left(  e_{(1)}{}^\mu \right)_{\rm H}  $,
\begin{equation}
\dot\sigma_ {\rm H}(2M)= -\lim_{r\to2M} \, \left[ \left(
 {\triangle\over r^2} \right)^2
{M\dot \Psi_{0\,\omega}  \over (1+4iM\omega)} \right] \nonumber
\end{equation}
(where $\dot\Psi_0$ is with respect to the Kinnersley tetrad). We
 thus obtain the
formula
\begin{equation}
{{\rm d}^2 E^{\rm hor}_\omega  \over {\rm d}t {\rm d}\Omega}
={1 \over 2^8 \pi  M^4
\left( 16 M^2 \omega^2 +1 \right) } \lim_{r\to2M} \,  \left[
\triangle^4
 \bigl| \dot\Psi_{0\,\omega}   \bigr|^2 \right]\quad
. \label{ehor}
\end{equation}

 At  spatial infinity the space-time becomes flat
 and one can therefore carry over standard results from the theory of
 gravity
linearized about Minkowski space, enabling one to derive the
formulae\cite{T2}
\begin{mathletters}
\label{e}
\begin{eqnarray}
 {{\rm d}^2 E^{\rm in}_\omega  \over {\rm d}t {\rm  d}\Omega} &&={1
 \over
64\pi\omega^2} \lim_{r\to\infty} \, \left[ r^2\bigl|
\dot\Psi_{0\,\omega}
 \bigr|^2\right] \label{e:1}\\
 {{\rm d}^2 E^{\rm out}_\omega  \over {\rm d}t {\rm d}\Omega} &&
={1 \over 4\pi\omega^2}
\lim_{r\to\infty} \, \left[ r^2 \bigl| \dot\Psi_{4\,\omega}
 \bigr|^2 \right]  \label{e:2}
\end{eqnarray}
\end{mathletters}
 for the incoming and outgoing energy flux  infinitely far from the
 black
hole. Formulae  (\ref{ehor}), (\ref{e})  demonstrate that
 $\bigl| \dot\Psi_0 \bigr|^2$ and
$\bigl| \dot\Psi_4 \bigr|^2$ directly
measure the energy flux of gravitational perturbations of a
Schwarzschild
black hole across the horizon and at infinity.  This physical
interpretation,
together with the other features already discussed, makes $\Psi_0$
and $\Psi_4$ the natural choice as field variables for the
semi-classical theory.

\section{REVIEW OF THE SEMI-CLASSICAL THEORY}
\label{semi-classical}

Teukolsky has provided us with a complete set of perturbed Weyl scalars
 $\dot\Psi_0$
and $\dot\Psi_4$ (henceforth collectively labelled $\dot\Psi_A$).
One could quantize the theory directly in terms of this set\cite{P},
however we choose to follow Candelas et al~\cite{CCH} by working in
terms of
the metric perturbation.  Thus we require a complete set of solutions
 for the metric perturbations $h_{\mu\nu}$. The relationship
between $\dot\Psi_A$ and $h_{\mu\nu}$ is obtained by perturbing the
defining
equations for the Weyl scalars. Consider for example
\begin{equation}
\Psi_0 = -C_{(1)(3)(1)(3)} = -C_{\mu\nu\rho\lambda} e_{(1)}{}^\mu
e_{(3)}{}^\nu
e_{(1)}{}^\rho e_{(3)}{}^\lambda
\end{equation}
 which when perturbed linearly becomes
\begin{eqnarray}
 \dot\Psi_0= &&-\dot{C}_{\mu\nu\rho\lambda} e_{(1)}{}^\mu e_{(3)}{}^\nu
e_{(1)}{}^\rho e_{(3)}{}^\lambda
 -C_{\mu\nu\rho\lambda} \left(\dot e_{(1)}{}^\mu e_{(3)}{}^\nu
 e_{(1)}{}^\rho
e_{(3)}{}^\lambda
 \right.    \nonumber \\ &&\left. +e_{(1)}{}^\mu \dot e_{(3)}{}^\nu
e_{(1)}{}^\rho
e_{(3)}{}^\lambda+e_{(1)}{}^\mu e_{(3)}{}^\nu \dot e_{(1)}
{}^\rho e_{(3)}{}^\lambda
+e_{(1)}{}^\mu e_{(3)}{}^\nu e_{(1)}{}^\rho \dot
e_{(3)}{}^\lambda\right).
\label{psi0Cexp}
\end{eqnarray}
Since the tetrad spans the tangent space we can write
$\dot{e}_{(a)}{}^\mu=A_{(a)} {}^{(b)} e_{(b)}{}^\mu$ for some functions
$A_{(a)}{}^{(b)}$; it follows that the  last four terms on the right
hand side
of (\ref{psi0Cexp}) vanish, as the only  non--zero, unperturbed
 Weyl scalar is $\Psi_2=-
C_{(1)(3)(4)(2)}$. In the first term one can use expressions provided
by
Barth and Christensen\cite{BC} for the  perturbed Riemann and Ricci
tensors to
express $\dot{C}_{\mu\nu\rho\lambda}$ in  terms of $h_{\mu\nu}$ and
(after some
tedious algebraic manipulation) arrive at the following formula for
 $\dot\Psi_0$ in
terms of  $h_{(a)(b)}$:
\begin{mathletters}
\label{psiNP}
\begin{equation}
\dot\Psi_0 = {1 \over 2} \left\{ (\delta+2\alpha)\delta h_{(1)(1)}
-(2D-3\rho) (\delta+2\alpha) h_{(1)(3)} +(D-
\rho) (D-\rho) h_{(3)(3)} \right\}. \label{psiNP:1}
\end{equation}
 A similar argument for $\dot\Psi_4$ yields
\begin{equation}
  \dot\Psi_4 = {1 \over 2} \left\{ (\delta^*+2\alpha) \delta^*
h_{(2)(2)} - (2\Delta+3\mu+4\gamma)
(\delta^*+2\alpha)  h_{(2)(4)} + (\Delta+\mu+2\gamma) (\Delta+\mu)
h_{(4)(4)} \right\} .  \label{psiNP:2}
\end{equation}
\end{mathletters}

Despite appearances Eqs. (\ref{psiNP}) can be inverted to yield
$h_{\mu\nu}$ in terms
of $\dot\Psi_A$, following a procedure due to Wald\cite{W} which is
 facilitated by a
choice of gauge reflecting the symmetry of the background space-time.
It is
then straightforward to obtain a complete set of solutions to the
equation of
motion for $h_{\mu\nu}$ from Teukolsky's complete set of Weyl
 scalars $\dot\Psi_A$.
Rather than repeat this lengthy derivation here however, we simply
state the
result (originally obtained by Chrzanowski\cite{C} using less direct
methods)
and verify that it does indeed constitute a complete set of solutions
to the
perturbed Einstein equations, by substituting into (\ref{psiNP}) and
reproducing Teukolsky's solution set (\ref{phis}).

We  write Chrzanowski's  complete, complex mode set as
\begin{equation}
 \left\{ h_{\mu\nu}^{\Lambda} (l,m,\omega, P;x),h_{\mu\nu}^{\Lambda*}
 (l,m,\omega, P;x) \right\}_{\Lambda,l,m,\omega, P}
\end{equation}
where $\Lambda\in\left\{{\rm in},{\rm up}\right\}$ and $P=\pm1$.
 The explicit form
of the modes for which $\Lambda={\rm in}$ is
\begin{mathletters}
\label{h}
\begin{equation}
 h_{\mu\nu}^{\rm in} (l,m,\omega, P;x)=N^{\rm in} \left\{
 \Theta_{\mu\nu}
 \,{}_{+2} Y_l^m(\theta,\phi) +P
\Theta_{\mu\nu}^* \,{}_{-2} Y_l^m(\theta,\phi) \right\} \,{}_{-2}
R_{l\omega}  ^{\rm in}(r)
 {\rm e}^{-i\omega  t}
\label{h:1}
\end{equation}
in the {\it ingoing radiation gauge} $h_{\mu\nu} e_{(1)}{}^\nu=0$,
$h_{\nu}{}^{\nu}=0$; when $\Lambda={\rm up}$ we have
\begin{equation}
 h_{\mu\nu}^{\rm up} (l,m,\omega, P;x) = N^{\rm up} \left\{
\Upsilon_{\mu\nu} \,{}_{-2} Y_l^m(\theta,\phi) +  P
\Upsilon_{\mu\nu}^* \,{}_{+2} Y_l^m(\theta,\phi) \right\} \,{}_{+2}
R_{l\omega}^{\rm up}(r) {\rm e}^{-i\omega  t}
\label{h:2}
\end{equation}
\end{mathletters}
 in the {\it outgoing radiation gauge} $h_{\mu\nu} e_{(2)}{}^\nu=0,$
$h_{\nu}{}^{\nu}=0$.  Here $N^\Lambda$ are constants which will be
fixed  by
the quantization  prescription, and  $\Theta_{\mu\nu}$,
$\Upsilon_{\mu\nu}$
are the  second-order differential operators:
\begin{eqnarray}
 \Theta_{\mu\nu}= &&-
e_{(1)}{}_\mu e_{(1)\nu}(\delta^*-2\alpha)(\delta^*-4\alpha)
-e_{(4)\mu}e_{(4)\nu}(D-
\rho)(D+3\rho) \nonumber \\
&&+{1 \over 2}\left(e_{(1)\mu}e_{(4)\nu}+e_{(4)\mu}e_{(1)\nu}\right)
\left[D(\delta^*-
4\alpha)+(\delta^*-4\alpha)(D+3\rho)\right]\nonumber \\
\Upsilon_{\mu\nu}= &&\rho^{-4}\left\{-
e_{(2)\mu}e_{(2)\nu}(\delta-2\alpha)(\delta-4\alpha)
-e_{(3)\mu}e_{(3)\nu}
(\Delta+5\mu-
2\gamma)(\Delta+\mu-4\gamma)\right.\nonumber \\ &&\left.
+{1 \over 2}\left(e_{(2)\mu}e_{(3)\nu}+e_{(3)\mu}e_{(2)\nu}\right)
\left[(\delta-
4\alpha)(\Delta+\mu-4\gamma)+(\Delta+4\mu-4\gamma)(\delta-4\alpha)
\right]\right\} .
 \label{thetaups}
\end{eqnarray}

The quantities $\,{}_{-2} R_{l\omega}  ^{\rm in}(r)$ and
 $\,{}_{+2} R_{l\omega}  ^{\rm up}(r)$
appearing in (\ref{h}) are those
particular solutions of the radial equation (\ref{req}) with $s=-2$ and
$s=+2$
respectively  which are specified by the boundary conditions
\begin{mathletters}
\label{rlims}
\begin{equation}
 \,{}_{-2} R_{l\omega}  ^{\rm in}(r) \sim \cases{ B_{l\omega}  ^{\rm
 in}
 \triangle^2 {\rm e}^{-i\omega  r_*} , & as $r\to2M$  \cr
\mathstrut &\cr r^{-1} {\rm e}^{-i\omega  r_*} +A_{l\omega}  ^{\rm in}
r^3
 {\rm e}^{+i \omega  r_*} ,& as $ r\to\infty $\cr } \quad,
\label{rlims:1}
\end{equation}
and
 \begin{equation}
 \,{}_{+2} R_{l\omega}  ^{\rm up}(r) \sim \cases{ A_{l\omega}  ^{\rm
 up}
 \triangle^{-2} {\rm e}^{-i\omega  r_*} + {\rm e}^{+i \omega  r_*},
& as $r\to2M $\cr
\mathstrut &\cr B_{l\omega}  ^{\rm up} r^{-5} {\rm e}^{+i \omega  r_*}
, &
as $r\to\infty$ \cr}
 \quad . \label{rlims:2}
\end{equation}
\end{mathletters}
 In the light of (\ref{simpleasymp}), (\ref{asymp2M}) it is clear
that $\,{}_{-2} R_{l\omega}  ^{\rm in}(r)$ and
$\,{}_{+2} R_{l\omega}  ^{\rm up}(r)$ are uniquely specified by the
above conditions; explicit
formulae for $A_{l\omega}  ^{\Lambda}$ and $B_{l\omega}  ^{\Lambda}$
will be produced in a later
section.

\begin{figure}
\centerline{\epsfig{figure=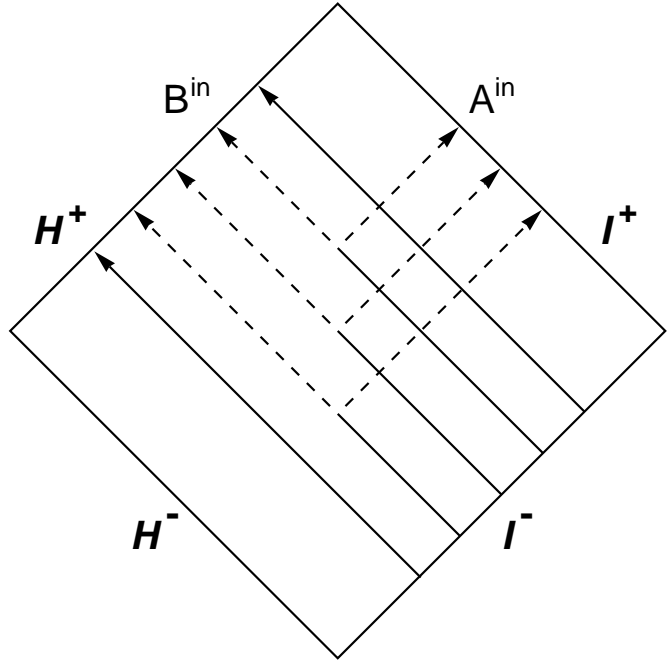,height=9truecm}}
\vskip1.5truecm
\centerline{\epsfig{figure=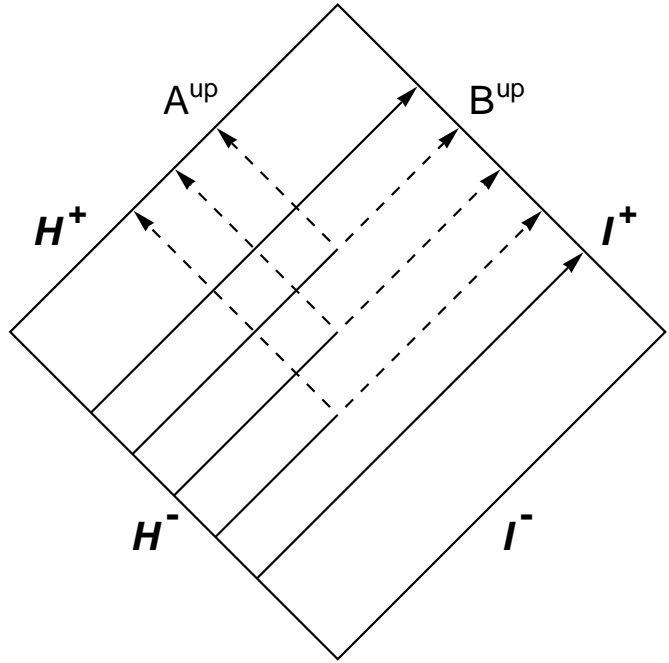,height=9truecm}}
\bigskip\caption{Representation of the `in' and `up' modes on a Penrose
diagram of the exterior Schwarzschild space-time. The `in' modes
emanate from ${\cal I}^-$ and are partial transmitted across ${\cal
H}^+$ and partly reflected out to ${\cal I}^+$. The `out' modes emanate from ${\cal H}^-$ and are partial transmitted out
to ${\cal I}^+$ and partly reflected across ${\cal H}^+$.}\label{modes}
\end{figure}

Use of the subscripts ``in'' and ``up'' derives from the analogy with
classical scattering demonstrated in the previous section; from
 (\ref{rlims:1}) we can
interpret  $\,{}_{-2} R_{l\omega}  ^{\rm in}(r) {\rm e}^{-i\omega  t}$
 as a unit-amplitude spherical wave
propagating {\it inwards} from infinity and being partially reflected
back out
to infinity and partially transmitted across the horizon, whereas from
 (\ref{rlims:2})
we  see that $\,{}_{+2} R_{l\omega}  ^{\rm up}(r) {\rm e}^{-i\omega
t}$
 represents a unit-amplitude
 spherical wave propagating  {\it upwards} from the past horizon, and
 being
partially reflected back and  partially transmitted out to infinity.
$|A_{l\omega}^{\Lambda}|^2$  is the    reflection coefficient and and
$|B_{l\omega}^{\Lambda}|^2$ is the transmission  coefficient
  for the scattering
process. The situation is depicted in Fig.~\ref{modes}.

The fact that the ``in'' and ``up'' modes (\ref{h}) are expressed in
  different
gauges will not cause  any difficulty later as we shall only use the
modes to
construct objects which are gauge  independent
 (namely $\bigl| \dot\Psi_A \bigr|^2$).

We now verify explicitly that Chrzanowski's mode set
comprises a complete set of solutions to the perturbed field equations
for
Schwarzschild space-time, and in the process derive a set of equations
which
will prove valuable for calculating expectation values of the perturbed
Weyl
scalars later.

First substitute (\ref{h:1}) for $h_{(a)(b)}$ in Eq. (\ref{psiNP:1}).
Since we are working in
the ingoing radiation gauge  for which $h_{(a)(1)}=0$, we have simply
\begin{eqnarray*}
  \dot\Psi_0 \left[ h^{\rm in} (l,m,\omega, P;x) \right] &&=
 {1 \over 2} (D-\rho) (D-\rho)
h_{(3)(3)} ^{\rm in}(l,m,\omega, P;x)  \\ =- {1 \over 2} N^{\rm in}
(D-\rho)
 && (D-\rho) (D-\rho)
(D+3\rho) \,{}_{+2} Y_l^m(\theta,\phi) \,{}_{-2} R_{l\omega}  ^{\rm
in}(r)
{\rm e}^{-i\omega  t}
\end{eqnarray*}
 where  (\ref{h:1}) and (\ref{thetaups}) have been used in obtaining
the second line.

Recalling that
\[
D= e_{(1)}{}^\mu \partial_\mu = {r^2\over\triangle}
 {\partial\over\partial{t}}  +
{\partial\over\partial{r}}
\]
and $\rho=-r^{-1}$, we see that $D\rho=\rho^2$ and consequently
\begin{eqnarray}
 \dot\Psi_0 \left[ h^{\rm in} (l,m,\omega, P;x) \right] && = -{1 \over
 2}
 N^{\rm in} DDDD
\,{}_{+2} Y_l^m \,{}_{-2} R_{l\omega}  ^{\rm in}(r) {\rm e}^{-i\omega
t}
  \nonumber \\ && = -{1 \over 2} N^{\rm in} \,{}_{+2}
  Y_l^m(\theta,\phi)
{\rm e}^{-i\omega  t} {\cal D}{\cal D}{\cal D}{\cal D} \,{}_{-2}
 R_{l\omega}^{\rm in}(r) \label{4Dform}
\end{eqnarray}
where
\begin{equation}
 {\cal D} = {{\rm d}\over {\rm d}{r}}  - {i\omega  r^2\over\triangle}
 .
\label{Ddef}
\end{equation}
 We now use the following result of Press and Teukolsky;\cite{PT}
 if $\,{}_{-2} R_{l\omega}  (r)$ is  any solution of the radial
 equation
 (\ref{req}) with $s=-2$,
then ${\cal D}{\cal D}{\cal D}{\cal D} \,{}_{-2} R_{l\omega}  (r)$ will
be a
 solution of (\ref{req})
with $s=+2$. So let  $\,{}_{+2} R_{l\omega}  ^{\rm in}(r)$ be that
particular
 solution of the $s=+2$
radial equation which is given by
\begin{equation}
 \,{}_{+2} R_{l\omega}  ^{\rm in}(r) = 4 {\cal D}{\cal D}{\cal
D}{\cal D} \,{}_{-2} R_{l\omega}  ^{\rm in}(r) \label{+2to-2}
\end{equation}
(the 4 is present to achieve
consistency with the normalizations used in Refs.~\cite{PT} and
\cite{C}).
 Then (\ref{4Dform}) becomes simply
\begin{equation}
\dot\Psi_0 \left[ h^{\rm in} (l,m,\omega, P;x) \right] = - {1\over8}
N^{\rm in}
 \,{}_{+2} R_{l\omega}  ^{\rm in}(r)
\,{}_{+2} Y_l^m(\theta,\phi) {\rm e}^{-i\omega  t}.
\end{equation}
In a similar manner one can substitute (\ref{h:1})  into Eq.
(\ref{psiNP:2})
 for $\dot\Psi_4$, and
also  (working in the outgoing gauge now) insert equation (\ref{h:2})
for the
``up'' mode into both (\ref{psiNP:1}) and (\ref{psiNP:2}). Altogether
one obtains \cite{BP}
\begin{mathletters}
\label{psieqs}
\begin{eqnarray}
  \dot \Psi_{0}^{\rm in} (l,m,\omega, P;x) && = -{1\over8} N^{\rm in}
\,{}_{+2} R_{l\omega} ^{\rm in}(r) \,{}_{+2} Y_l^m(\theta,\phi) {\rm
e}^{-i\omega t} \label{psieqs:1}\\ \dot \Psi_{4}^{\rm in} (l,m,\omega,
P;x) && = -{1\over 8 r^4} N^{\rm in} \left( { {\cal R}\hbox{e} } \,
C_{l\omega} + 12iM\omega P \right) \,{}_{-2} R_{l\omega} ^{\rm in}(r)
\,{}_{-2} Y_l^m(\theta,\phi) {\rm e}^{-i\omega t}
\label{psieqs:2}\\ \dot
\Psi_{0}^{\rm up} (l,m,\omega, P;x) && = -{1\over8} N^{\rm up} \left( {
{\cal R}\hbox{e} } \, C_{l\omega} -12iM\omega P \right) \,{}_{+2}
R_{l\omega} ^{\rm up}(r) \,{}_{+2} Y_l^m(\theta,\phi) {\rm e}^{-i\omega
t}
\label{psieqs:3}\\ \dot \Psi_{4}^{\rm up} (l,m,\omega, P;x) && =
-{1\over
8 r^4} N^{\rm up} \left| C_{l\omega} \right|^2 \,{}_{-2} R_{l\omega}
^{\rm
up}(r) \,{}_{-2} Y_l^m(\theta,\phi) {\rm e}^{- i\omega t}
 \label{psieqs:4} \end{eqnarray} \end{mathletters} where we are using
 the
obvious notation $ \dot \Psi_{A}^{\Lambda} (l,m,\omega, P;x)$ for $
\dot{\Psi}_A \left[ h^\Lambda (l,m,\omega, P;x) \right]$
 and
 \begin{equation}
 C_{l\omega} = (l-1)l(l+1)(l+2) + 12iM\omega  \quad .
 \end{equation}
 $\,{}_{-2} R_{l\omega}  ^{\rm up}(r)$ is defined by the equation
\begin{equation}
 \,{}_{-2} R_{l\omega} ^{\rm up}(r) = {\triangle^2 \over 4 \left|
C_{l\omega} \right|^2} {\cal D}^\dagger {\cal D}^\dagger {\cal
D}^\dagger
{\cal D}^\dagger \left( \triangle^2 \,{}_{+2} R_{l\omega} ^{\rm up}(r)
\right) \label{4Ddaggerform}
 \end{equation}
and  is known\cite{PT} to be a solution of the $s=-2$ radial equation.
We stress that since $\dot\Psi_A$ is gauge independent, the final
expressions
(\ref{psieqs}) do not depend on the particular gauges chosen to
facilitate their derivation.

Eq.(\ref{psieqs}) is clearly in perfect agreement with Teukolsky's
solution (\ref{phis}), and we conclude that the mode set (\ref{h}) does
indeed constitute a complete set of solutions to the equation of motion
for the metric perturbation.

\begin{figure}
\centerline{\epsfig{figure=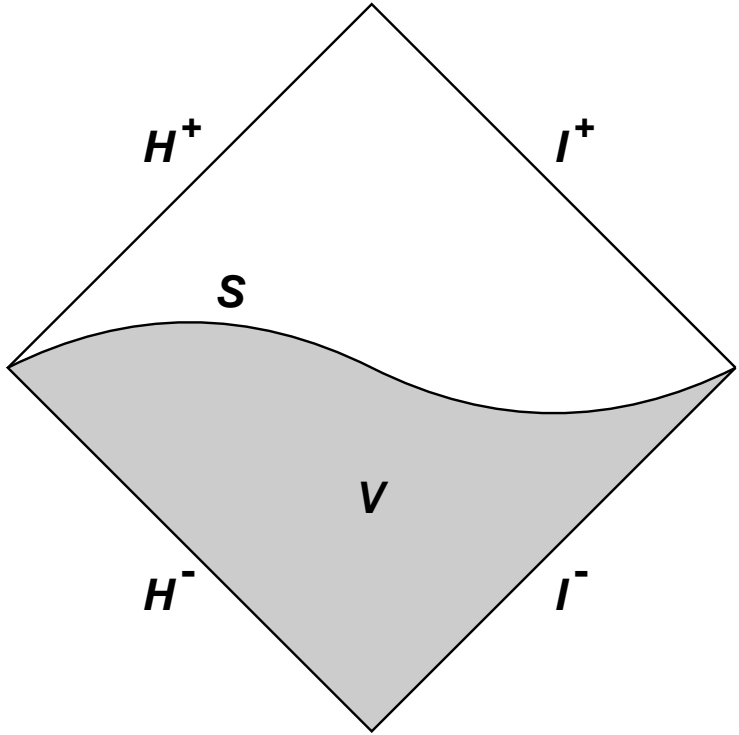,height=10truecm}}
\bigskip\caption{ The space-time region ${\cal V}$ of the exterior
Schwarzschild
space-time.}\label{spacetime}
\end{figure}

The constants $N^\Lambda$ appearing in (\ref{h}) are fixed in the
quantum
theory by demanding that the mode set satisfy the othonormality
conditions
\begin{equation}
\langle {1 \over \sqrt{16 \pi}} h^\Lambda(l,m,\omega, P;x) ,
  {1 \over \sqrt{16 \pi}}h^{\Lambda'}(l',m',\omega',P';x) \rangle =
 \delta_{\Lambda\Lambda'} \delta_{ll'} \delta_{mm'}
 \delta(\omega-\omega')
\delta_{PP'} \label{on} \end{equation} where the inner product
$\langle\quad,\quad\rangle$ is defined as follows for arbitrary complex
symmetric tensor fields $\psi_{\alpha\beta}$ and $\phi_{\alpha\beta}$:
\begin{equation}
  \langle\psi,\phi\rangle ={i\over2} \int_{\cal S}{\rm d}\Sigma^\mu
\left(\psi^{\alpha\beta*}\nabla_\mu\tilde\phi_{\alpha\beta}-
\phi^{\alpha\beta}\nabla_\mu\tilde\psi _{\alpha\beta}^*
+2\tilde\phi_{\alpha\mu}\nabla_\beta\tilde\psi^{\alpha\beta*}-
2\tilde\psi_{\alpha\mu}^* \nabla_\beta\tilde\phi^{\alpha\beta}\right)
\end{equation}
 ($\tilde\psi_{\alpha\beta}$, $\tilde\phi_{\alpha\beta}$ are the
trace-free parts of $\psi_{\alpha\beta}$, $\phi_{\alpha\beta}$
respectively).  ${\cal S}$ is an arbitrary spacelike hypersurface in
the
exterior Schwarzschild space-time, so one can divide the left hand side
of
(\ref{on}) into three integrals
\[
 \int_{\cal S}=\int_{\tilde{\cal S}}-\int_{{\cal H}^-}-\int_{{\cal
 I}^-}
\]
where $\tilde{\cal S}$ is the
surface enclosing the volume ${\cal V}$ shown in Fig. \ref{spacetime}.

 The integral on the closed hypersurface $\tilde{\cal S}$ converts to
 an
integral over the space-time region ${\cal V}$ via Stokes' theorem;
this
subsequently vanishes because its integrand is identically zero by
virtue
of the linearized field equations (to which $h_{\mu\nu}^{\Lambda}
(l,m,\omega, P;x)$ are solutions). The left hand side of (\ref{on})
thus
splits conveniently into just two integrals over ${{\cal H}^-}$ and
${{\cal I}^-}$, each of which can be evaluated explicitly since one
knows
the form of the radial functions $\,{}_{-2} R_{l\omega} ^{\rm in}$,
$\,{}_{+2} R_{l\omega} ^{\rm up}$ at both the horizon and spatial
infinity
(see (\ref{rlims})). To proceed rigorously one transforms $(t,r)$ to
Kruskal null co-ordinates
\[
 U  = -{\rm e}^{(r_*-t)/(4M)} , \qquad V  =  {\rm e}^{(r_*+t)/(4M)}
\]
 and employs
\begin{eqnarray*}
\left( {\rm d}\Sigma^\mu \right)_{{\cal H}^-} &&=-e_{(2)}{}^\mu
{4Mr^2\over U}
{\rm d}U \sin\theta {\rm d}\theta {\rm d}\phi  \\ \left( {\rm d}
\Sigma^\mu
\right)_{{\cal I}^-}
 &&= e_{(1)}{}^\mu {2M\triangle\over V} {\rm d}V \sin\theta {\rm
 d}\theta
{\rm d}\phi \end{eqnarray*}
 as future-directed surface elements for the null hypersurfaces ${{\cal
I}^-}$ and ${{\cal H}^-}$ respectively. Integrations over angular
co-ordinates can be performed using the orthonormality relations
(\ref{B3}) for the spherical harmonics. One discovers that
Chrzanowski's
mode set satisfies the orthonormality conditions (\ref{on}) provided
\begin{mathletters} \label{norms} \begin{eqnarray}
 \left| N^{\rm in} \right|^2 &&={1\over 4\omega^5} \label{norms:1} \\
\left| N^{\rm up} \right|^2 &&={16\over (2M)^5 p_\omega }
\label{norms:2}
\end{eqnarray} \end{mathletters} where \begin{equation} p_\omega = 2M
\omega \left( 1 +4 M^2 \omega^2 \right) \left( 1 + 16 M^2 \omega^2
\right)
\quad .  \label{pomega} \end{equation} These normalization factors
differ
from those given in Ref. \cite{CCH} only in that we are working in
units
where $G=1$ while the authors of Ref. \cite{CCH} chose units where
$16\pi
G=1$.

 An arbitrary linear perturbation of the Schwarzschild background can
 be
expanded as follows in terms of the complex mode set (\ref{h}):
\begin{equation}
 h_{\mu\nu}(x) = \sum_K
\left\{ a_K h_{\mu\nu}^K (x) + a_K^*  h_{\mu\nu}^{K*}(x) \right\}
\label{hmunu}
\end{equation}
 where $K$ is a shorthand for $(\Lambda,l,m,\omega, P)$. When the
 theory is
quantized $h_{\mu\nu}$, $a_K$ and $a_K^*$ become operators on the
Hilbert
space of quantum states of the system, subject to canonical commutation
relations which (since the mode set is orthonormal) take the  simple
form
\begin{equation}
\left[ { a}_K,{ a}_{K'}^\dagger \right] = \delta_{KK'} ,\qquad \left[ {
a}_K,{ a}_{K'}         \right] = 0 =\left[ { a}_K^\dagger,{
a}_{K'}^\dagger
\right]  . \label{commrels}
\end{equation}
 Consider now the expectation value $\langle B|\bigl| \dot\Psi_A
\bigr|^2|B\rangle$, where $|B\rangle$ is the vacuum state associated
with
the mode set (\ref{h}), called the {\it Boulware vacuum}.  Since
$\dot\Psi_A$ is linear in $h_{\mu\nu}$ and its derivatives (see
(\ref{psiNP})), one can substitute expansion (\ref{hmunu}) for
$h_{\mu\nu}$ and obtain \begin{eqnarray*}
 \langle B| \bigl| \dot\Psi_A \bigr|^2 |B\rangle \equiv \langle B| &&
\sum_K \left\{ { a}_K \dot\Psi_A \left[ h^K \right] +{ a}_K^\dagger
\dot\Psi_A \left[ h^{K*} \right] \right\} \times \\ && \qquad \sum_{K'}
\left\{ { a}_{K'} \dot\Psi_A^* \left[ h^{K'} \right] +{ a}_{K'}^\dagger
\dot\Psi_A^* \left[ h^{K'*} \right] \right\} |B\rangle \\ &&=\sum_K
\left|
\dot\Psi_A \left[ h^K \right] \right|^2 \end{eqnarray*}
 Written out fully this is
\begin{equation}
 \langle B| \bigl| \dot\Psi_A \bigr|^2 |B\rangle = \sum_{l=2}^\infty
\sum_{m=-l}^{+l} \int_0^\infty {\rm d}\omega \> \sum_{P=\pm 1} \left\{
\bigl| \dot \Psi_{A}^{\rm in} (l,m,\omega, P;x) \bigr|^2 +\bigl| \dot
\Psi_{A}^{\rm up} (l,m,\omega, P;x) \bigr|^2 \right\} \label{Bmodesum}
\end{equation} where $\dot \Psi_{A}^{\Lambda} (l,m,\omega, P;x)$ for
$\Lambda={\rm in},{\rm up}$ are given explicitly by (\ref{psieqs}).

Although the Boulware state $|B\rangle$ corresponds to the usual notion
of a
vacuum at large radii, it becomes pathological as one approaches the
horizon, in the sense  that  the components of the renormalized
expectation value of the energy-momentum tensor for the scalar  and
electromagnetic fields in the Boulware vacuum diverge in a
freely-falling frame as $r\to2M$ (Ref.~\cite{CF}; see also \cite{Ca},
\cite{JMO}). It is thus only appropriate for studying a body whose
radius exceeds $2M$ such as a neutron star.

The {\it Unruh vacuum} $| U \rangle$ is the vacuum state of a mode set
whose ``in''
 modes are positive frequency with respect to the co-ordinate $t$
(i.e.~they are precisely the $h_{\mu\nu}^{\rm in} (l,m,\omega, P;x)$
defined by (\ref{h:1})), but whose ``up'' modes are defined to be
positive
frequency with respect to the Kruskal null co-ordinate $U$. The
renormalized energy-momentum tensor for the scalar field in this state
is
regular on ${{\cal H}^+}$ but not on ${{\cal H}^-}$, and corresponds to
a
flux of black body radiation as $r\to\infty$. The Unruh vacuum
approximates the state of the field long after the gravitational
collapse
of a massive body.

Finally the {\it Hartle--Hawking vacuum} $| H \rangle$, defined to be
that
of a mode set whose ``in'' modes are positive frequency with respect to
$V$ and whose ``up'' modes are positive frequency with respect to $U$,
yields a renormalized energy-momentum tensor which is regular on both
horizons, but one then has a bath of thermal ``Hawking'' radiation at
infinity, so that $| H \rangle$ is not a true vacuum state in the usual
sense. It corresponds instead to a black hole in unstable equilibrium
with
an infinite bath of thermal radiation.

If the calculation leading to (\ref{Bmodesum}) is performed using the
Unruh and Hartle-Hawking mode sets instead of the Boulware mode set
(\ref{h}), one obtains \cite{CCH} \begin{eqnarray}
 \langle U| \bigl| \dot\Psi_A \bigr|^2 | U \rangle &&=
 \sum_{l=2}^\infty
\sum_{m=-l}^{+l} \int_0^\infty {\rm d}\omega \> \sum_{P=\pm 1} \times
\nonumber \\ && \times \left\{ \bigl| \dot \Psi_{A}^{\rm in}
(l,m,\omega,
P;x) \bigr|^2 +\coth \left( 4 \pi M \omega \right) \bigl| \dot
\Psi_{A}^{\rm up} (l,m,\omega, P;x) \bigr|^2 \right\} \label{Umodesum}
\end{eqnarray}
 and
\begin{eqnarray}
 \langle H| \bigl| \dot\Psi_A \bigr|^2 | H \rangle &&=
 \sum_{l=2}^\infty
\sum_{m=-l}^{+l} \int_0^\infty {\rm d}\omega \> \sum_{P=\pm 1} \coth
\left( 4 \pi M \omega \right) \times \nonumber \\ &&\times \left\{
\bigl|
\dot \Psi_{A}^{\rm in} (l,m,\omega, P;x) \bigr|^2 +\bigl| \dot
\Psi_{A}^{\rm up} (l,m,\omega, P;x) \bigr|^2 \right\} \label{Hmodesum}
\end{eqnarray}
 respectively. On account of their distributional character the product
 of
two field operators $h_{(a)(b)}$ (or their derivatives) evaluated at
the
same space-time point is a priori ill-defined; as a result the
expectation
value of any such product is infinite. Therefore, since $\bigl|
\dot\Psi_A
\bigr|^2$ is quadratic in $h_{\mu\nu}$ and its derivatives (see
(\ref{psiNP})), expressions (\ref{Bmodesum}--\ref{Hmodesum}) are all
infinite. Rather than become imbrued in the construction of a
renormalization scheme for $\langle\bigl| \dot\Psi_A \bigr|^2\rangle$
analogous to that employed in the evaluation of $\langle
T^{\mu\nu}\rangle$ for quantum fields of lower spin, we elect instead
to
work with the {\it differences} \begin{mathletters} \label{modediff}
\begin{eqnarray}
  \langle \bigl| \dot\Psi_A \bigr|^2 \rangle ^{H-U} &&\equiv \langle H|
\bigl| \dot\Psi_A \bigr|^2 | H \rangle - \langle U| \bigl| \dot\Psi_A
\bigr|^2 | U \rangle
		      \nonumber \\ &&= 2 \sum_{l=2}^\infty
\sum_{m=-l}^{+l} \int_0^\infty {\rm d}\omega \> \sum_{P=\pm 1} { \bigl|
\dot \Psi_{A}^{\rm in} (l,m,\omega, P;x) \bigr|^2 \over \left({\rm
e}^{8\pi M\omega} -1\right) } \label{modediff:1} \end{eqnarray} and
\begin{eqnarray}
  \langle \bigl| \dot\Psi_A \bigr|^2 \rangle ^{U-B} &&\equiv \langle U|
\bigl| \dot\Psi_A \bigr|^2 | U \rangle - \langle B| \bigl| \dot\Psi_A
\bigr|^2 |B\rangle \nonumber \\ &&= 2 \sum_{l=2}^\infty
\sum_{m=-l}^{+l}
\int_0^\infty {\rm d}\omega \> \sum_{P=\pm 1} { \bigl| \dot
\Psi_{A}^{\rm
up} (l,m,\omega, P;x) \bigr|^2
      \over \left({\rm e}^{8\pi M\omega}   -1\right)
} \quad .  \label{modediff:2}
\end{eqnarray}
\end{mathletters}
These differences are automatically finite for $r>2M$, by virtue of the
fact that the divergent component of the expectation value of such a
two-point product is purely geometrical and hence state-independent.

Using these differences we can, of course, work out the difference
\[
 \langle \bigl| \dot\Psi_A \bigr|^2 \rangle ^{H-B} \equiv \langle H|
\bigl| \dot\Psi_A \bigr|^2 | H \rangle - \langle B| \bigl| \dot\Psi_A
\bigr|^2 |B\rangle =
  \langle \bigl| \dot\Psi_A \bigr|^2 \rangle ^{H-U} + \langle \bigl|
\dot\Psi_A \bigr|^2 \rangle ^{U-B} . \] It is a consequence of the
time-reversal invariance of this difference and the transformation
properties of the Kinnersley tetrad under time-reversal that
\begin{equation}
  (r-2M)^4 \langle \bigl| \dot\Psi_0 \bigr|^2 \rangle ^{H-B} =
	   16 r^4 \langle \bigl| \dot\Psi_4 \bigr|^2 \rangle ^{H-B} .
\label{04relation} \end{equation}

Substituting (\ref{psieqs}) for $\dot \Psi_{A}^{\Lambda} (l,m,\omega,
P;x)$ in (\ref{modediff}) we obtain finally \begin{mathletters}
\label{diff} \begin{eqnarray}
 &&\langle \bigl| \dot\Psi_0 \bigr|^2 \rangle^{H-U} = {1\over 2^{8}
 \pi}
\sum_{l=2}^\infty (2l+1) \int_0^\infty{{\rm d}\omega \over \omega^5}
{\left| \,{}_{+2} R_{l\omega} ^{\rm in}(r) \right|^2 \over \left({\rm
e}^{8\pi M\omega} -1\right)} \label{diff:1}\\ &&\langle \bigl|
\dot\Psi_4
\bigr|^2 \rangle^{H-U} = {1 \over 2^{8} \pi r^8} \sum_{l=2}^\infty
(2l+1)
\int_0^\infty { {\rm d}\omega \over \omega^5 } {\left| C_{l\omega}
\right|^2 \left| \,{}_{-2} R_{l\omega} ^{\rm in}(r) \right|^2 \over
\left({\rm e}^{8\pi M\omega} -1\right)} \label{diff:2}\\ &&\langle
\bigl|
\dot\Psi_0 \bigr|^2 \rangle^{U-B} = {1 \over 4 \pi (2M)^5}
\sum_{l=2}^\infty (2l+1) \int_0^\infty { {\rm d}\omega \over p_\omega }
{\left| C_{l\omega} \right|^2 \left| \,{}_{+2} R_{l\omega} ^{\rm up}(r)
\right|^2 \over \left({\rm e}^{8\pi M\omega} -1\right)} \label{diff:3}
\\
&&\langle \bigl| \dot\Psi_4 \bigr|^2 \rangle^{U-B} = {1 \over 4 \pi
(2M)^5
r^8} \sum_{l=2}^\infty (2l+1) \int_0^\infty { {\rm d}\omega \over
p_\omega
} {\left| C_{l\omega} \right|^4 \left| \,{}_{-2} R_{l\omega} ^{\rm
up}(r)
\right|^2 \over \left({\rm e}^{8\pi M\omega} -1\right)} \label{diff:4}
\end{eqnarray} \end{mathletters} where the addition theorem (\ref{B4})
for
spin--weighted spherical harmonics has been employed to perform the sum
over $m$, and (\ref{norms}) has been substituted for $|N^\Lambda|^2$.

\section{ POWER SERIES REPRESENTATION FOR $\,{}_{\lowercase{s}}
R_{\lowercase{l}\omega}  $}
\label{powerseries}

In order to evaluate the expressions (\ref{diff}), an explicit
representation of the radial function $\,{}_{s} R_{l\omega} $ valid
throughout the region exterior to the black hole is required. In what
follows we use the methods of Leaver\cite{L} to construct a general
solution of the spin $s$ radial equation (\ref{req}) in terms of a
convergent power series about $r=2M$.  This procedure has much better
stability properties than performing a straight numerical integration
of
the differential equation (\ref{req}).

In light of (\ref{asymp2M}) one particular solution of the spin $s$
radial
equation (\ref{req}) can be expanded as \begin{equation}
 \,{}_{s} R_{l\omega} (r) = \triangle^{-s} {\rm e}^{-i\omega r_*}
 \,{}_{s}
S_{l\omega} (r) \label{Rs} \end{equation} where \begin{equation}
 \,{}_{s} S_{l\omega} (r) = \sum_{k=0}^\infty a_k(l,\omega, s)
 \left(1-{2M
\over r} \right)^k .\label{Ss} \end{equation}
 The coefficients $a_k(l,\omega, s)$ are determined by substituting
(\ref{Rs}) into the radial equation (\ref{req}) and changing variable
from
$r$ to $x= 1-(2M/r)$. This yields the following differential equation
for
$\,{}_{s} S_{l\omega}$:  \begin{eqnarray}
 \left[ x(1-x)^2 {{\rm d}^2\over {\rm d}{x}^2} +\left\{ (1-s)
\left(1-x^2\right) -2x(1-
 x) -4iM\omega \right\} {{\rm d}\over {\rm d}{x}} \right. \nonumber &&
 \\
\left. + \left\{ (2s-1) {4iM\omega\over(1-x)} +s(s-1) -l(l+1) \right\}
\right] \,{}_{s} S_{l\omega} && = 0.  \label{Sdiffeq} \end{eqnarray}
 Substitution of expansion (\ref{Rs}) for $\,{}_{s} S_{l\omega} $ then
yields the following three term recursion relation which determines the
$a_k(l,\omega, s)$:  \begin{eqnarray}
 && \quad k(k-4iM\omega-s)a_k \nonumber \\ &&+\left[ -3(k-1)^2
+(k-1)(s+4iM\omega) + 4iM\omega(2s-1) +s(s-1) -l(l+1) \right] a_{k-1}
\nonumber \\ && + \left[ 3(k-2)^2 +s(k-2) - s(s-1) +l(l+1) \right]
a_{k-2}
+\left[ -(k-3)^2 -s(k-3) \right] a_{k-3} = 0
 \label{recrel}
\end{eqnarray}
with the initial conditions $a_0 =1$, $a_{-1}=a_{-2}=0$.

Another independent solution of the Teukolsky equation is
(c.f.~Eq.~(\ref{asymp2M})) \begin{equation} \,{}_{s} R_{l\omega} (r) =
{\rm e}^{+i \omega r_*} \,{}_s\tilde{S}_{l\omega} (r) \label{Stilde}
\end{equation}
 with \begin{equation} {}_s\tilde{S}_{l\omega} (r) = \sum_{k=0}^\infty
\tilde{a}_k(l,\omega, s) \left(1-{2M \over r} \right)^k \end{equation}
 say.  Substitution of (\ref{Stilde}) into (\ref{req}) yields the
 complex
conjugate of (\ref{Sdiffeq}) with $s$ replaced by $-s$, so that
\begin{equation}
 {}_s\tilde{S}_{l\omega}   = \,{}_{-s} S_{l\omega}  ^* \quad.
\end{equation}
 One concludes that the general solution of the spin $s$ radial
 equation
(\ref{req}) is \begin{equation}
 \qquad \,{}_{s} R_{l\omega} (r) = c_1 \triangle^{-s} {\rm e}^{-i\omega
r_*} \,{}_{s} S_{l\omega} (r) +c_2 {\rm e}^{+i \omega r_*} \,{}_{-s}
S_{l\omega} ^*(r) \label{gensol} \end{equation}
 where $\,{}_{s} S_{l\omega}  (r)$ is given by (\ref{Rs}), the
 coefficients
$a_k(l,\omega, s)$ being determined by the recursion relation
(\ref{recrel}).

The choice of coefficients $c_1 $, $c_2 $ which yields either
$\,{}_{-2}
R_{l\omega} ^{\rm in}(r)$ or $\,{}_{+2} R_{l\omega} ^{\rm up}(r)$ is
determined by comparing the form of either of these particular
solutions
as $r\to2M$ (see Eq.~(\ref{rlims})) with that of the general solution
(\ref{gensol}). One deduces the representations \begin{mathletters}
\label{StoR} \begin{eqnarray}
  \,{}_{-2} R_{l\omega} ^{\rm in} &&= B_{l\omega} ^{\rm in} \triangle^2
{\rm e}^{-i\omega r_*} \,{}_{-2} S_{l\omega} \label{StoR:1}
\\ \,{}_{+2}
R_{l\omega} ^{\rm up} &&= A_{l\omega} ^{\rm up} \triangle^{-2} {\rm
e}^{-i\omega r_*} \,{}_{+2} S_{l\omega} + {\rm e}^{+i \omega r_*}
\,{}_{-2} S_{l\omega} ^* \label{StoR:2} \end{eqnarray} which are valid
throughout the exterior region.

The corresponding representations of $\,{}_{+2} R_{l\omega} ^{\rm in}$
and
$\,{}_{-2} R_{l\omega} ^{\rm up}$ are obtained by substituting
(\ref{StoR:1}) and (\ref{StoR:2}) into (\ref{+2to-2}) and
(\ref{4Ddaggerform}) respectively, and performing the derivative
operations explicitly; derivatives of $\,{}_{s} S_{l\omega} $ of order
$\ge2$ can be eliminated with the aid of (\ref{Sdiffeq}). The result is
\begin{eqnarray}
 \,{}_{+2} R_{l\omega} ^{\rm in} &&= 4 B_{l\omega} ^{\rm in}
\triangle^{-2} {\rm e}^{-i\omega r_*} \left\{ \alpha_{l\omega}
\,{}_{-2}
S_{l\omega} +\beta_{l\omega} \triangle \,{}_{-2} S_{l\omega} ' \right\}
		   \label{StoR:3} \\
 4 \left| C_{l\omega} \right|^2 \,{}_{-2} R_{l\omega} ^{\rm up} &&=
A_{l\omega} ^{\rm up} \triangle^{-1} {\rm e}^{-i\omega r_*} \left\{
\gamma_{l\omega} \,{}_{+2} S_{l\omega} ' +\delta_{l\omega} \,{}_{+2}
S_{l\omega} \right\} \nonumber \\
 &&\qquad +{\rm e}^{+i \omega r_*} \left\{ \gamma_{l\omega} \triangle
{\,{}_{-2} S_{l\omega} ^*}' +\epsilon_{l\omega} \,{}_{-2} S_{l\omega}
^*
\right\} \label{StoR:4} \end{eqnarray} \end{mathletters} where the
prime
denotes ${\rm d}/{\rm d}r$ and \begin{eqnarray*}
 \alpha_{l\omega} (r)= && (l-1)l(l+1)(l+2) \triangle^2 - 2i \omega r^2
\left\{ (l- 1)(l+2) r \left(4r^2 -10Mr +4M^2 \right) \right. \\
 && \left. +10Mr^2 -40 M^2 r +24M^3 \right\} -
 4 \omega^2 r^4 \left\{ 10Mr -24M^2 +3 (l-1)(l+2) \triangle \right\} \\
&& + 8 i \omega^3 r^6 \{r+2M\} +16 \omega^4 r^8 \\
 \beta_{l\omega} (r)= &&-2 i \omega r^2 \left( 4Mr - 12M^2 +2
 (l-1)(l+2)
\triangle \right) + 8 i \omega^3 r^6 \\ \gamma_{l\omega} (r)= &&2 i
\omega
r^2 \left\{ 4 \left( 2r^2 -3Mr -3M^2 \right) +2(l-2)(l+3)\triangle
\right\} - 8 i \omega^3 r^6 \\ \delta_{l\omega} (r)= &&
(l-1)l(l+1)(l+2)
\triangle -
   4 i \omega  r \left\{ 8r^2 -9Mr -6M^2 + (l-2)(l+3) r(2r-3M) \right\}
   \\
 &&- 4 \omega^2 r^3 \left\{ 2(2r+3M) + (l-2)(l+3) r
\right\} +24 i \omega^3 r^5 \\
\epsilon_{l\omega}  (r)= && (l-1)l(l+1)(l+2) \triangle^2 \\ &&+2i
\omega  r^2
\left\{ 2 \left( 8r^3 -15Mr^2 -12M^2r +12M^3 \right)
 +2 (l-2)(l+3) (2r-M) \triangle \right\}  \\
 &&- 4 \omega^2 r^4 \left\{ 2 \left( 6r^2 -7Mr
-12M^2 \right) +3 (l-2)(l+3) \triangle \right\} \\
 &&-8 i \omega^3 r^6 (r+2M) + 16\omega^4 r^8
\quad .
\end{eqnarray*}

 It is reassuring to check that in the limit as $r\to2M$ these latter
representations reduce to
\begin{mathletters}
\label{newasymp}
\begin{eqnarray}
 \,{}_{+2} R_{l\omega}  ^{\rm in} && \sim {16 i  (2M)^4 p_\omega  \over
 (1 + 2iM\omega)}
		B_{l\omega}  ^{\rm in} \triangle^{-2} {\rm
		e}^{-i\omega  r_*}    \label{newasymp:1}\\
 \,{}_{-2} R_{l\omega}  ^{\rm up} && \sim - {i (1 + 2i M\omega)  \over
 16  (2M)^4 p_\omega}   A_{l\omega}  ^{\rm up}
 \triangle^2 {\rm e}^{-i\omega  r_*}  - {i  (2M)^4 p_\omega  \over
 \left| C_{l\omega} \right|^2(1-2iM\omega) } {\rm e}^{+i \omega  r_*}
\label{newasymp:2}
\end{eqnarray}
which is consistent with their expected behavior (\ref{asymp2M}) in
this limit.
Their behavior as $r \to \infty$ is derived in Appendix
\ref{asymptoticapp}. For completeness we give the leading behavior here
\begin{eqnarray}
\,{}_{+2} R_{l\omega}  ^{\rm in} &&\sim
64 \omega^4 {{\rm e}^{-i\omega  r_*} \over r}
+{|C_{l\omega}|^2 \over 4 \omega^4} A_{l\omega}  ^{\rm in}
{{\rm e}^{+i \omega  r_*}\over r^5}
\label{newasymp:3} \\
\,{}_{-2} R_{l\omega}  ^{\rm up} &&\sim {4 \omega^4 \over \left|
C_{l\omega} \right|^2} B_{l\omega}  ^{\rm up} r^3 {\rm e}^{+i \omega
r_*}
\label{newasymp:4}
\end{eqnarray}
\end{mathletters}

One can now insert the above power series representations (\ref{StoR})
for the radial
functions into the expressions (\ref{diff}) for the differences in
expectation
 values of the perturbed Weyl scalars, thereby obtaining
\begin{mathletters}
\label{Sdiff}
\begin{eqnarray}
&&\langle  \bigl| \dot\Psi_0 \bigr|^2 \rangle^{H-U} = {1 \over 2^4 \pi
\triangle^4 } \sum_{l=2}^\infty (2l+1)
\int_0^\infty {{\rm d} \omega  \over \omega^5} {|B_{l\omega}  ^{\rm
in}|^2 \over \left({\rm e}^{8\pi M\omega}   -1\right)} \left|
\alpha_{l\omega}   \,{}_{-2} S_{l\omega}   +\beta_{l\omega}   \triangle
\,{}_{-2} S_{l\omega}  ' \right|^2  \label{Sdiff:1}\\
&&\langle \bigl| \dot\Psi_4 \bigr|^2 \rangle^{H-U} =  {\triangle^4
\over 2^{8} \pi r^8} \sum_{l=2}^\infty (2l+1)
\int_0^\infty {{\rm d}\omega  \over \omega^5} {|B_{l\omega}  ^{\rm
in}|^2 \over \left({\rm e}^{8\pi M\omega}   -1\right)} |C_{l\omega}|^2
|\,{}_{-2} S_{l\omega}  |^2    \label{Sdiff:2}\\
&&\langle  \bigl| \dot\Psi_0 \bigr|^2 \rangle^{U-B} = {1\over
4 \pi (2M)^5} \sum_{l=2}^\infty (2l+1) \int_0^\infty {{\rm d}\omega
|C_{l\omega}|^2 \over p_\omega  \left({\rm e}^{8\pi M\omega}
-1\right)}
\nonumber \\ &&\qquad\quad\times \left\{ \left[|A_{l\omega}  ^{\rm
up}|^2 \triangle^{-4} |\,{}_{+2} S_{l\omega}  |^2
+|\,{}_{-2} S_{l\omega}  |^2 \right] +2{ {\cal R}\hbox{e} } \left[
A_{l\omega}  ^{\rm up} e^{-2i\omega  r_*} \triangle^{-2} \,{}_{+2}
S_{l\omega}
\,{}_{-2} S_{l\omega}   \right] \right\}       \label{Sdiff:3}\\
&&\langle  \bigl| \dot\Psi_4 \bigr|^2 \rangle^{U-B} =
 {1\over 2^{6} \pi r^8 (2M)^5} \sum_{l=2}^\infty (2l+1) \int_0^\infty
 {{\rm d}\omega  \over p_\omega
\left({\rm e}^{8\pi M\omega}   -1\right)} \nonumber
\\ &&\qquad\quad\times \left\{ \left[ |A_{l\omega}  ^{\rm up}|^2
\triangle^{-2} \left|
\gamma_{l\omega}
 \,{}_{+2} S_{l\omega}  ' +\delta_{l\omega}   \,{}_{+2} S_{l\omega}
 \right|^2 + \left|\gamma_{l\omega}  ^* \triangle \,{}_{-2}
 S_{l\omega}  '
+\epsilon_{l\omega}  ^*
 \,{}_{-2} S_{l\omega}   \right|^2 \right]\right. \nonumber
 \\ &&\qquad\quad\left. + 2{ {\cal R}\hbox{e} } \left[
A_{l\omega}  ^{\rm up} e^{-2i\omega  r_*} \triangle^{-1} \left(
\gamma_{l\omega}   \,{}_{+2} S_{l\omega}  ' +\delta_{l\omega}
\,{}_{+2} S_{l\omega}
\right) \left( \gamma_{l\omega}  ^* \triangle \,{}_{-2} S_{l\omega}  '
+\epsilon_{l\omega}  ^* \,{}_{-2} S_{l\omega}   \right) \right]
\right\} \quad . \label{Sdiff:4}
\end{eqnarray}
\end{mathletters}

\section{THE REFLECTION AND TRANSMISSION COEFFICIENTS}
\label{ABcoefficients}
It remains to derive formulae from which $ B_{l\omega}  ^{\rm in}$ and
$ A_{l\omega}  ^{\rm up}$ can be
determined explicitly;  expressions (\ref{Sdiff}) for the expectation
values can then be evaluated numerically.

Such a formula for $B_{l\omega}  ^{\rm in}$ is derived by comparing the
power series
representation for $\,{}_{-2} R_{l\omega}  ^{\rm in}$ (which is valid
for all $r>2M$) with its
asymptotic expansion valid only at large radii (asymptotic expansions
of the various radial functions are derived in Appendix
\ref{asymptoticapp}). Thus from (\ref{StoR:1}) and (\ref{C1}) we have
at large $r$
\begin{eqnarray}
B_{l\omega}  ^{\rm in}  {\rm e}^{-i\omega  r_*} \,{}_{-2} S_{l\omega}
(r)
=&& {\rm e}^{-i\omega  r_*} \left[{1\over r^5} +{\cal O} \left( {1\over
r^6} \right)\right] \nonumber \\
&&+A_{l\omega}  ^{\rm in} {\rm e}^{+i \omega  r_*} \left[ {1\over r}+
{e_1\over r^2} +{e_2\over r^3}
+{e_3\over r^4} +{e_4\over r^5}
+{\cal O} \left( {1\over r^6} \right) \right]
\label{larger}
\end{eqnarray}
with
\begin{eqnarray*}
e_1 &&= b_1 +2(2M) \\
e_2 &&= b_2 +2(2M)b_1 +3(2M)^2 \\
e_3 &&= b_3 +2(2M)b_2 +3(2M)^2 b_1 +4(2M)^3 \\
e_4 &&= b_4 +2(2M)b_3 +3(2M)^2 b_2 +4(2M)^3 b_1 +5(2M)^4
\end{eqnarray*}
where the $b_i$ are given in Appendix \ref{asymptoticapp}.
By ignoring terms of order $r^{-5}$ on the right hand side of this
equation one could write down an approximate formula for the ratio
$A_{l\omega}  ^{\rm in}/B_{l\omega}  ^{\rm in}$, evaluate it
numerically at some suitably large value of $r$,
and hence determine $|B_{l\omega}  ^{\rm in}|^2$ from the Wronskian
relation
\begin{equation}
\left| B_{l\omega}  ^{\rm in} \right|^2 =
4 \omega^5 \left\{ (2M)^5 p_\omega  +
{|C_{l\omega}|^2 \over 2^6 \omega^3} \left| {A_{l\omega}  ^{\rm
in}\over B_{l\omega}  ^{\rm in}} \right|^2
\right\}^{-1},
\label{wronskian}
\end{equation}
this last equation being a variant on one of a complete set of
Wronskian
relations  between the reflection and transmission
coefficients which are derived in Appendix \ref{relations}.
However a more accurate approximation can be obtained by first applying
the
operator ${\cal D}^\dagger$ (recall (\ref{Ddef})) to both sides of
(\ref{larger}), so that
one need only ignore
terms of order $r^{-6}$ to arrive at the approximation
\begin{equation}
{A_{l\omega}  ^{\rm in}\over B_{l\omega}  ^{\rm in}} \approx
 \left[ {f_1\over r} +{f_2\over r^2}+{f_3\over r^3}+{f_4\over r^4}
+{f_5\over r^5}  \right]^{-1}
{\rm e}^{-2i\omega  r_*} {{\rm d}\over {\rm d}{r}}  \left(\,{}_{-2}
S_{l\omega}   \right)
\label{AoverB}
\end{equation}
for the ratio of the incoming coefficients, where
\begin{eqnarray*}
f_1 &&= 2i\omega  \\
f_2 &&= 2i\omega  \left[e_1 +(2M) \right] -1 \\
f_3 &&= 2i\omega  \left[e_2 +(2M)e_1 +(2M)^2 \right]
-2 e_1 \\
f_4 &&= 2i\omega  \left[ e_3 +(2M)e_2 +(2M)^2 e_1 +(2M)^3 \right]
-3 e_2\\
f_5 &&= 2i\omega  \left[e_4 +(2M)e_3 +(2M)^2 e_2 +(2M)^3 e_1 +(2M)^4
\right]
-4 e_3 \quad .
\end{eqnarray*}
In practice the right hand side of (\ref{AoverB}) is evaluated for
large and increasing
values of $r$ until it has converged to the desired accuracy; the
result is
then inserted in (\ref{wronskian}) to yield $|B_{l\omega}  ^{\rm
in}|^2$.

A slightly different procedure is used to evaluate the reflection
amplitude
$A_{l\omega}  ^{\rm up}$. A comparison of the power series expansion
(\ref{StoR:2}) for $\,{}_{+2} R_{l\omega}  ^{\rm up}$ with
its asymptotic form (\ref{C3}) at large radii yields
\begin{equation}
A_{l\omega}  ^{\rm up} {\rm e}^{-i\omega  r_*} \,{}_{+2} S_{l\omega}
+ \triangle^2 {\rm e}^{+i \omega  r_*} \,{}_{-2} S_{l\omega}  ^*
=
B_{l\omega}  ^{\rm up} {\rm e}^{+i \omega  r_*} \left[ {1\over r}
+{g_2\over r^2} +{g_3\over r^3}
+{g_4\over r^4}+{g_5\over r^5}
+{\cal O} \left({1\over r^6}\right) \right]
\label{Aupeq}
\end{equation}
with
\begin{eqnarray*}
g_2 &&= c_1 -4M \\
g_3 &&= c_2 -4Mc_1 +4M^2 \\
g_4 &&= c_3 -4Mc_2 +4M^2 c_1 \\
g_5 &&= c_4 -4Mc_3 +4M^2 c_2
\end{eqnarray*}
where the $c_i$ are given in Appendix \ref{asymptoticapp},
which when operated on by ${\cal D}$ gives
\begin{eqnarray*}
A_{l\omega}  ^{\rm up} {\rm e}^{-i\omega  r_*}
&&\left[ {{{\rm d}\over {\rm d}{r}}  \left(\,{}_{+2} S_{l\omega}
\right)}
-{2i\omega  r^2 \over \triangle} \,{}_{+2} S_{l\omega}   \right]
+\triangle {\rm e}^{+i \omega  r_*} \left[ \triangle {{{\rm d}\over
{\rm d}{r}}  \left(\,{}_{-2} S_{l\omega}  ^* \right)}
+4(r-M) \,{}_{-2} S_{l\omega}  ^* \right] \\
&&=
-B_{l\omega}  ^{\rm up} {\rm e}^{+i \omega  r_*} \left[ {1\over r^2}
+{2 g_2 \over r^3}
+{3 g_3 \over r^4}+{4 g_4 \over r^5}+{5 g_5 \over r^6}
+{\cal O} \left({1\over r^7}\right) \right] \quad .
\end{eqnarray*}
To this last equation one now adds (\ref{Aupeq}) multiplied by
\begin{equation}
h(r)=\left[ {1\over r} +{h_2\over r^2} +{h_3\over r^3}
+{h_4\over r^4}+{h_5\over r^5} \right]
\end{equation}
where
\begin{eqnarray*}
h_2 &&=g_2 \\
h_3 &&=2g_3 -g_2 h_2 \\
h_4 &&=3g_4 -g_3 h_2 -g_2 h_3 \\
h_5 &&=4g_5 -g_4 h_2 -g_3 h_3 -g_2 h_4
\end{eqnarray*}
so that the right hand side of the resulting equation is ${\cal O}
\left({
r^{-7}}\right)$ and can be ignored at large radii. An accurate
approximation for
$A_{l\omega}  ^{\rm up}$ is therefore given by
\begin{eqnarray}
A_{l\omega}  ^{\rm up} \approx
-\triangle^2 {\rm e}^{2i\omega  r_*}
&&\left[ {{{\rm d}\over {\rm d}{r}}  \left( \,{}_{-2} S_{l\omega}  ^*
\right)}
+ \left\{ h(r) +{4(r-M)\over\triangle} \right\} \,{}_{-2} S_{l\omega}
^* \right] \times \nonumber \\
\times
&&\left[  {{{\rm d}\over {\rm d}{r}}  \left(\,{}_{+2} S_{l\omega}
\right)}
+\left\{ h(r) -{2i\omega  r^2 \over \triangle} \right\} \,{}_{+2}
S_{l\omega}   \right]^{-1}
\label{Aupapprox}
\end{eqnarray}
which is evaluated in the same way as (\ref{AoverB}).

\begin{figure}
\centerline{\epsfig{figure=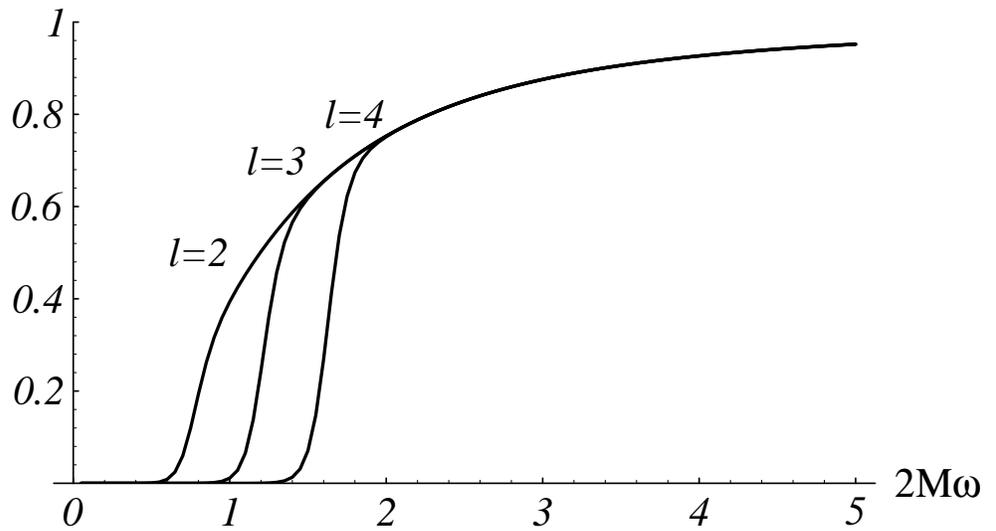,height=10truecm}}
\bigskip\caption{$(2M)^{10}  |B_{l\omega}  ^{\rm in}|^2$   as a
function of
$2M\omega$  for $l=2,3,4$.}\label{Bin}
\end{figure}

\begin{figure}
\centerline{\epsfig{figure=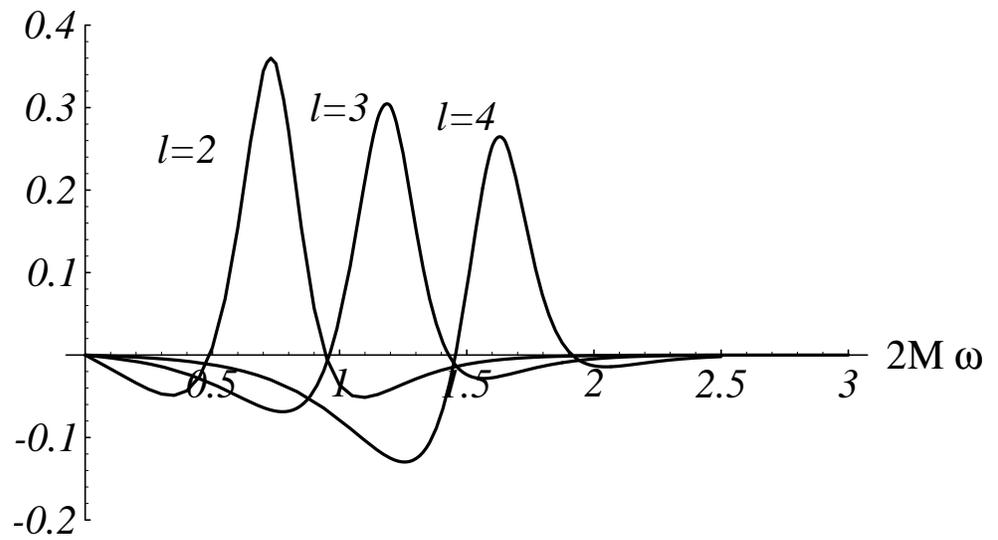,height=10truecm}}
\bigskip\caption{ The imaginary part of $(2M)^{-4} A_{l\omega}^{\rm
up}$ as a function of
$2M\omega$  for $l=2,3,4$.}\label{Aup}
\end{figure}

\begin{figure}
\centerline{\epsfig{figure=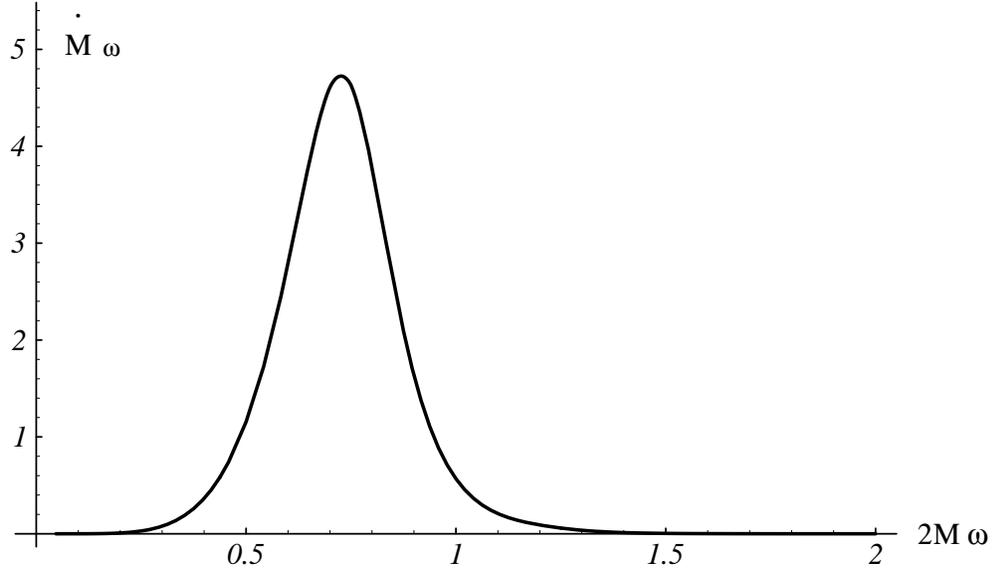,height=9.5truecm}}
\bigskip\caption{${\dot M}_\omega \times 10^5$, the luminosity spectrum
due to graviton emission from a Schwarzschild black
hole.}\label{spectrum}
\end{figure}

The values of the reflection and transmission amplitudes, which we have
computed numerically, are displayed graphically in Figs. \ref{Bin}
through \ref{Aup}.  We have verified that these values are
correct using the following (independent) checks:

\noindent
(i) The rate of decrease of mass of the black hole due to graviton
emission is
given by the formula~\cite{P}
\begin{equation}
\dot M = \int_0^\infty {\rm d}\omega \>  \dot M_{\omega}
\end{equation}
where
\begin{equation}
\dot M_{\omega}   = \sum_{l=2}^\infty \sum_{m=-l}^{+l} \sum_{P=\pm 1}
{\omega \over 2\pi} {1\over \left({\rm e}^{8\pi M\omega}   -1\right)}
{ {\rm d} E^{\rm hor}_\omega  / {\rm d}t \over {\rm d} E^{\rm
in}_\omega  / {\rm d}t }
\quad .
\end{equation}
Substituting first (\ref{ehor}), (\ref{e:1}) for the energy fluxes, and
then
$\dot \Psi_{0}^{\rm in} (l,m,\omega, P;x)$ (see Eq. (\ref{psieqs})) for
$\dot\Psi_0$, one obtains the following
expression for the fractional absorption of incoming radiation by the
black
hole (spherical harmonics are eliminated with the aid of (\ref{B3})):
\begin{eqnarray*}
{ {\rm d} E^{\rm hor}_\omega  / {\rm d}t \over {\rm d} E^{\rm
in}_\omega  / {\rm d}t }
&&=
{1\over (2M)^4}
{4 \omega^2 \over \left(16 M^2 \omega^2 +1\right)}
{  \left[ \triangle^4 | \,{}_{+2} R_{l\omega}  ^{\rm in} |^2
\right]_{r=2M}
\over
 \left[ r^2 \left| \,{}_{+2} R_{l\omega}  ^{\rm in} \right|^2
 \right]_{r\to\infty} }  \\
&&=
{(2M)^5 \over 4 \omega^5} p_\omega  | B_{l\omega}  ^{\rm in} |^2
\end{eqnarray*}
where the last line follows from (\ref{newasymp:1}) and (\ref{C2}), and
hence
\begin{equation}
\dot M_{\omega}   = {(2M)^5 \over 4 \pi} \sum_{l=2}^\infty (2l+1)
     { p_\omega  | B_{l\omega}  ^{\rm in} | ^2  \over
	   \omega^4 \left({\rm e}^{8\pi M\omega}   -1\right)}
\quad .  \label{mdot}
\end{equation}
Using the
values obtained numerically for $| B_{l\omega}  ^{\rm in} |^2 $ in the
above equations one obtains the luminosity spectrum shown in Fig.
\ref{spectrum} and the
value $\dot M = ( 3{\cdot}84  \times 10^{-6} )M^{-2}$ for the total
luminosity due to
graviton emission; both are in close agreement with the previous
results of
Page\cite{P} (Page has $\dot M = ( 3{\cdot}81  \times 10^{-6} )M^{-2}$;
we believe
the slight discrepancy to be caused by his choice of an upper limit for
the
$\omega$  integral which is not high enough to ensure 3 significant
figure accuracy.)
\noindent
(ii) When $2M\omega\ll1$ the spin $s$ radial equation can be solved
analytically in two overlapping regions, one of which includes the
horizon, and
the other of which extends to infinity; solutions are expressed in
terms of
ordinary and confluent hypergeometric functions in the
respective regions (see Ref.~\cite{CS}). By matching these general
solutions in the
region of overlap, and then specialising to the particular solutions
$\,{}_{-2} R_{l\omega}  ^{\rm in}$, $\,{}_{+2} R_{l\omega}  ^{\rm up}$
in turn,
one can derive analytic approximations for the reflection and
transmission amplitudes which are valid in the limit as
$2M\omega\to0$.
The derivation is performed in detail in Appendix E. The exact
numerical values
for the amplitudes are consistent with the analytic approximation in
the
limit $2M\omega\to0$.
\noindent (iii) A final check on the numerical
values we have obtained for $| B_{l\omega}  ^{\rm in} |^2$ and
$\left|A_{l\omega}  ^{\rm up}\right|^2$ is provided by
the Wronskian relations derived in Appendix D. From (\ref{D1}) and
(\ref{D3}) the following
relation between the `in' and `up'  reflection coefficients may be
obtained:
\begin{equation}
 |A_{l\omega}  ^{\rm in} |^2 = { 16 \omega^7 \over (2M)^9 \left( 16 M^2
 \omega^2 +1 \right)
p_\omega  } |A_{l\omega}  ^{\rm up} |^2 \quad .
\end{equation}
Substituting this relation into (\ref{D2}) then yields
\begin{equation}
{ \left( 4M^2\omega^2 +1 \right) \left|C_{l\omega}\right|^2 \over
16 (2M)^8 p_\omega^2} \left|A_{l\omega}  ^{\rm up}\right|^2
+{ (2M)^5 p_\omega  \over 4 \omega^5} | B_{l\omega}  ^{\rm in} |^2 = 1
\quad .
\label{check}
\end{equation}
We have shown our numerical results to be consistent with the above
equation.

\section{ASYMPTOTIC ANALYSIS}
\label{asymptotics}
Before we discuss the numerical evaluation of Eqs. (\ref{diff}),
it is useful to consider their asymptotic behavior in the limits
$r\to 2M$ and $r \to \infty$.

Consider first the limit $r\to 2M$.  The leading behavior
of the various radial functions in the limit as $r\to2M$ may be deduced
from
Eqs~(\ref{StoR}), and (\ref{newasymp}); Eqs. (\ref{diff}) then give
\begin{mathletters}
\label{2M}
\begin{eqnarray}
&&\langle \bigl| \dot\Psi_0 \bigr|^2 \rangle^{H-U}
\sim {(2M)^9  \over  \pi \triangle^4} \sum_{l=2}^\infty (2l+1)
\int_0^\infty {\rm d}\omega
{\left( 1+16 M^2 \omega^2\right) p_\omega   | B_{l\omega}  ^{\rm in}
|^2 \over
 \omega^4  \left({\rm e}^{8\pi M\omega}   -1\right)} \label{2M:1}
\\
&&\langle \bigl| \dot\Psi_4 \bigr|^2 \rangle^{H-U}
\sim {\triangle^4 \over 2^{8} \pi  (2M)^8} \sum_{l=2}^\infty (2l+1)
\int_0^\infty {\rm d}\omega  {  |C_{l\omega} |^2  | B_{l\omega}  ^{\rm
in}  |^2 \over \omega^5  \left({\rm e}^{8\pi M\omega}   -1\right)}
\label{2M:2}
\end{eqnarray}
\end{mathletters}
The integrals in Eq.~(\ref{2M}) may be evaluated numerically to give:
\begin{mathletters}
\begin{eqnarray}
 \lim_{r \to 2M} \> (1 - 2M/r)^4\langle \bigl| \dot\Psi_0 \bigr|^2
 \rangle^{H-U}
       &&\approx 1{\cdot}96 \times 10^{-4} \> (2M)^{-6}\\
 \lim_{r \to 2M} \>  (1 - r/2M)^{-4}\langle \bigl| \dot\Psi_4 \bigr|^2
 \rangle^{H-U}
      && \approx  9{\cdot}14 \times 10^{-5} \> (2M)^{-6}
\end{eqnarray}
\end{mathletters}

The same method cannot be used to evaluate the limit
of $\langle \bigl| \dot\Psi_A \bigr|^2 \rangle^{U-B} $ as $r\to 2M$.
For writing
\begin{equation}
\langle \bigl| \dot\Psi_A \bigr|^2\rangle=\sum_{l=2}^\infty
(2l+1)\psi_{A\,l}
\end{equation}
one finds (by substituting (\ref{C2}) in (\ref{diff:1})) that
\begin{mathletters}
\label{2Mprob}
\begin{equation}
\Delta^4 \psi_{0\,l}^{U-B} \sim {11 (2M)^2 \over 2^{4} 15 \pi   } -
		       \Delta^4 \psi_{0\,l}^{H-U}
\label{2Mprob:1}
\end{equation}
and
\begin{equation}
\psi_{4\,l}^{U-B} \sim {11\over 2^{8} 15 \pi} \, (2M)^{-6} .
\label{2Mprob:2}
\end{equation}
\end{mathletters}
In neither case will the sum over $l$ converge since the terms approach
a constant times $(2l+1)$ for large $l$ (note from (\ref{2Mprob:1})
that we have
already seen that the corresponding sum over $\psi_{0\,l}^{H-U}$
converges).  The problem is that the limit as $r \to 2M$ and  the  sum
over $l$ do not commute in this case since our asymptotic expansions
for the Teukolsky functions are not {\it uniform} in $l$.  Nevertheless
Eqs. (\ref{2Mprob}) do provide a useful check against the numerical
results.

To find the asymptotic behavior of  $\langle \bigl| \dot\Psi_0 \bigr|^2
\rangle^{U-B} $ and $\langle \bigl| \dot\Psi_4 \bigr|^2 \rangle^{U-B} $
as $r\to
2M$ we need asymptotic expressions for the Teukolsky functions which
are uniform in $l$. To this end we follow Candelas \cite{Ca}.
Near $r=2M$ the Teukolsky equation (\ref{req}) may be approximated by
the equation
\begin{equation}
  \left[(r-2M) {{\rm d}^2\over {\rm d}{r}^2} + (s+1) {{\rm d}\over {\rm
  d}{r}}  +{2M \omega  (2M\omega  - is)
   \over (r- 2M)}  -{(l-s)(l+s+1) \over 2M}\right] \,{}_{s}
   R_{l\omega}  (r) = 0. \label{reqapprox}
\end{equation}
Defining $\displaystyle \xi = (r/2M - 1)^{1/2}$ this becomes
\begin{equation}
  \left[{{\rm d}^2\over {\rm d}{\xi}^2} + {(2s+1) \over \xi} {{\rm
  d}\over {\rm d}{\xi}}  + {8M \omega  (2M\omega  - is)
   \over \xi^2}  - 4 (l-s)(l+s+1) \right] \,{}_{s} R_{l\omega}  (r) =
   0, \label{reqapproxxi}
\end{equation}
which admits solutions in terms of modified Bessel functions. Since it
is clear from
Eq. (\ref{2Mprob}) that the asymptotic forms
of $\langle \bigl| \dot\Psi_0 \bigr|^2
\rangle^{U-B} $ and $\langle \bigl| \dot\Psi_4 \bigr|^2 \rangle^{U-B} $
as $r\to
2M$ are determined by the contribution from large $l$, we may further
approximate
the constant term in the potential in Eq. (\ref{reqapproxxi})   by $-4
l^2$ and then the
solutions are given by
\[
  \xi^{-s} K_{s+4iM\omega}  (2l\xi)  \qquad {\rm and } \qquad \xi^{-s}
  I_{-s-4iM\omega}  (2l\xi).
\]
These solutions are uniformly valid for large $l$.

    We now concentrate on $\langle \bigl| \dot\Psi_0 \bigr|^2
    \rangle^{U-B} $ and write
\begin{equation}
  \,{}_{+2} R_{l\omega}  ^{\rm up} \sim \alpha_l  \, \xi^{-2}
  K_{2+4iM\omega}  (2l\xi) + \beta_l \, \xi^{-2} I_{-2-4iM\omega}
  (2l\xi).
	     \label{besselform}
\end{equation}
Following the arguments of Candelas, as $l \to \infty$ for fixed $\xi$,
$\,{}_{+2} R_{l\omega}  ^{\rm up} (\xi) \to 0$ and so $\beta_l$ is an
exponentially small function of
$l$ that  will not therefore contribute to the leading asymptotic
behavior of  $\langle \bigl| \dot\Psi_0 \bigr|^2 \rangle^{U-B} $ as $r
\to 2M$.
Furthermore  comparison of (\ref{besselform}) with (\ref{rlims:2}) for
fixed $l$ as $\xi \to 0$ yields
\begin{equation}
   \alpha_l = {2 i \Gamma(3+4iM\omega) \sinh (4\pi M \omega) {\rm e}^{2
   i M \omega}
		  \over  \pi l ^{2 + 4 iM \omega}  } .
\end{equation}
The elementary identity
\[
    \left| \Gamma(3+4iM\omega) \right|^2 = {8 \pi p_\omega  \over
    \sinh(4\pi M \omega)}
\]
then enables us to write
\begin{equation}
    |\alpha_l|^2 = {32 p_\omega  \sinh (4 \pi M\omega) \over \pi l^4}
    .
\end{equation}
Since for large $l$,  $|C_{l\omega}|^2 \sim l^8$, it follows that to
leading order
\begin{eqnarray}
  \sum_{l=2}^\infty (2l+1) \left| C_{l\omega} \right|^2 \left|
  \,{}_{+2} R_{l\omega}  ^{\rm up}(r) \right|^2  &\sim&
  {64 p_\omega  \sinh (4 \pi M \omega) \over \xi^4}
	    \int\limits_0^\infty {\rm d}l \> l^5 \left|K_{2+4iM\omega}
	    (2l\xi)\right|^2 \nonumber \\
    &=& {64 p_\omega^2 \over 5 \xi^{10}},
\end{eqnarray}
where we have used Eq. (6.576.3) of \cite{GR}.  Inserting this into Eq.
(\ref{diff:3})
we obtain the leading asymptotic form as $r \to 2M$ as
\begin{eqnarray}
\Delta^4 \langle \bigl| \dot\Psi_0 \bigr|^2 \rangle^{U-B} &\sim& {16
(2M)^4  \over 5 \pi (r - 2M)}
\int_0^\infty  {\rm d}\omega  {p_\omega  \over \left({\rm e}^{8\pi
M\omega}   -1\right)}  \nonumber \\
  &=& {191 \over 2^4 \, 315 \pi} {(2M)^3 \over (r-2M)}
  \label{hardasymp} \\
  &\approx& 1{\cdot}21 \times 10^{-2} {(2M)^3 \over (r-2M)} , \nonumber
\end{eqnarray}
or equivalently,
\begin{equation}
\lim_{r \to 2M} \> (r/2M-1)^5 \langle \bigl| \dot\Psi_0 \bigr|^2
\rangle^{U-B} \approx
	     1{\cdot}21 \times 10^{-2} \> (2M)^{-6}   .
\end{equation}
Since, by (\ref{2M:1}), $\Delta^4 \langle \bigl| \dot\Psi_0 \bigr|^2
\rangle^{H-U} $ is finite on the horizon it follows that   the
asymptotic form of $\Delta^4 \langle \bigl| \dot\Psi_0 \bigr|^2
\rangle^{H-B}$ as $r \to 2M$ is also given by
(\ref{hardasymp}).  Then from (\ref{04relation})
\begin{eqnarray}
 \langle \bigl| \dot\Psi_4 \bigr|^2 \rangle^{H-B}  &\sim&
       {191 \over 2^{8} \, 315 \pi} \, {1 \over (2M)^5  (r-2M)}
       \label{hardasymp2} \\
  &\approx& 7{\cdot}54 \times 10^{-4} \> {1 \over (2M)^5 (r-2M)} .
  \nonumber
\end{eqnarray}
Since,   by (\ref{2M:2}), $\langle \bigl| \dot\Psi_4 \bigr|^2
\rangle^{H-U} $ vanishes at the horizon
it follows that the asymptotic form of $\langle \bigl| \dot\Psi_4
\bigr|^2
 \rangle^{U-B} $ as $r \to 2M$ is also given by (\ref{hardasymp2}).

We now turn to the asymptotic forms at infinity.
Substituting the asymptotic expansions (\ref{C3}) and (\ref{C4}) for
$\,{}_{+2} R_{l\omega}  ^{\rm up}$ and $\,{}_{-2} R_{l\omega}  ^{\rm
up}$ respectively in Eqs.  (\ref{diff:3}),
(\ref{diff:4}) yields the following formulae which describe the leading
behavior of $\langle \bigl| \dot\Psi_A \bigr|^2 \rangle^{U-B}$ in the
limit as
$r\to\infty$:
\begin{mathletters}
\label{inf}
\begin{eqnarray}
\langle \bigl| \dot\Psi_0 \bigr|^2 \rangle^{U-B}
&&\sim
{1 \over 4 \pi (2M)^5 r^{10}}
\sum_{l=2}^\infty (2l+1)
\int_0^\infty {\rm d}\omega
{  |C_{l\omega} |^2  | B_{l\omega}  ^{\rm up}  |^2 \over
p_\omega  \left({\rm e}^{8\pi M\omega}   -1\right)}
\label{inf:1}\\
\langle \bigl| \dot\Psi_4 \bigr|^2 \rangle^{U-B}
&&\sim
{4 \over  \pi (2M)^5 r^{2}}
\sum_{l=2}^\infty (2l+1)
\int_0^\infty {\rm d}\omega
{ \omega^8  | B_{l\omega}  ^{\rm up}  |^2 \over
p_\omega  \left({\rm e}^{8\pi M\omega}   -1\right)} \quad .
\label{inf:2}
\end{eqnarray}
Using the Wronskian relation (\ref{D1}), the last equation may be
rewritten as
\begin{equation}
\langle \bigl| \dot\Psi_4 \bigr|^2 \rangle^{U-B} \sim
{(2M)^5 \over 4 \pi  r^{2}}
\sum_{l=2}^\infty (2l+1)
\int_0^\infty {\rm d}\omega
{   p_\omega  | B_{l\omega}  ^{\rm in}  |^2 \over
\omega^2 \left({\rm e}^{8\pi M\omega}   -1\right)} \quad \label{inf:3}
\end{equation}
\end{mathletters}
  Eq. (\ref{inf:3})
is in accord with Eq. (\ref{mdot}) for the luminosity since Eq.
(\ref{e:2})  implies
that for $r \to \infty$
\begin{equation}
 \langle \bigl| \dot\Psi_{4\,\omega}   \bigr|^2 \rangle^{U-B} \sim
 {\omega^2  \dot M_\omega  \over r^2}.
\end{equation}
The integrals in Eq.~(\ref{inf}) may be evaluated numerically to give:
\begin{mathletters}
\begin{eqnarray}
 \lim_{r \to \infty} \> (r/2M)^{10} \langle \bigl| \dot\Psi_0 \bigr|^2
 \rangle^{U-B}
      && \approx 2{\cdot}91 \times 10^{-2}\>(2M)^{-6}\\
 \lim_{r \to \infty}  \> (r/2M)^2 \langle \bigl| \dot\Psi_4 \bigr|^2
 \rangle^{U-B}
      && \approx  8{\cdot}42 \times 10^{-6}\>(2M)^{-6}
\end{eqnarray}
\end{mathletters}

Finally, we consider  the  limit as $r\to\infty$ of
\[
\langle \bigl| \dot\Psi_A \bigr|^2 \rangle^{H-B}=
\langle \bigl| \dot\Psi_A \bigr|^2 \rangle^{H-U} + \langle \bigl|
\dot\Psi_A \bigr|^2 \rangle^{U-B} .
\]
We find that that
\begin{equation}
\psi_{0\,l}^{H-B}
\sim
16 \psi_{4\,l}^{H-B}
\sim
{1\over 2^{4} (2M)^4 15  \pi  r^2 } ,
\end{equation}
where Wronskian relations (\ref{D1}) and  (\ref{D2}) have been used.
As before these forms provide a useful check on our numerical results
but are not good enough to yield the asymptotic forms.

The simplest way to obtain the asymptotic forms at infinity is to use
the fact that at infinity the calculation reduces to a  problem in
Minkowski space.  Then quantum effects are negligible (except in
providing the appropriate temperature) and we may tak

e over the standard classical results given by Eq. (\ref{e}).  Firstly,
using spherical symmetry, we have
\[
   \bigl| \dot\Psi_{0\,\omega}   \bigr|^2 = - 64 \pi \omega^2
   T_\omega^{\rm in}{}_r{}^t \qquad {\rm and} \qquad
      \bigl| \dot\Psi_{4\,\omega}   \bigr|^2 = 4 \pi \omega^2
      T_\omega^{\rm out}{}_r{}^t
\]
where $T_\omega^{\rm in}{}_r{}^t$ and
$ T_\omega^{\rm out}{}_r{}^t$ denote the energy flux components of
frequency $\omega$  in the energy-momentum
tensors associated with the incoming and outgoing
gravitational waves, respectively. Then, since we are dealing with a
thermal bath
of gravitons at the Hawking temperature, $1/(8\pi M)$, we have
\begin{eqnarray*}
   - \langle  T_\omega^{\rm in}{}_r{}^t \rangle =
	\langle     T_\omega^{\rm out}{}_r{}^t \rangle &&=
		  2 \times \int\limits_{k_z > 0} {{\rm d}^3 {\bf k}
		  \over (2\pi)^3} \>
			{k_z\> \delta(k-\omega)  \over \left({\rm
			e}^{8\pi M\omega}   -1\right)}\\
	&&= 2\times {1 \over 8 \pi^2} {\omega^3 \over \left({\rm
	e}^{8\pi M\omega}   -1\right)}
\end{eqnarray*}
where the factor of $2$ arises from the number of polarization states,
and
multiplies  the ingoing and outgoing energy flux for a real massless
scalar
field in a thermal state with the Hawking temperature in Minkowski
space.
Thus, as $r \to \infty$ we have
\begin{mathletters}
\label{HBinf}
\begin{eqnarray}
\langle \bigl| \dot\Psi_0 \bigr|^2 \rangle^{H-B} &&\sim
16 \langle \bigl| \dot\Psi_4 \bigr|^2 \rangle^{H-B}
 \sim  {16 \over  \pi }
\int_0^\infty {\rm d}\omega  {   \omega^5 \over \left({\rm e}^{8\pi
M\omega}   -1\right)}\\
    &&= {1 \over 2016 \pi} \,  (2M)^{-6} \\
    && \approx 1{\cdot}58 \times 10 ^{-4} \> (2M)^{-6}
\end{eqnarray}
\end{mathletters}
Since, by Eq. (\ref{inf}) $\langle \bigl| \dot\Psi_0 \bigr|^2
\rangle^{U-B}$ and $\langle \bigl| \dot\Psi_4 \bigr|^2 \rangle^{U-B}$
vanish at infinity, Eq. (\ref{HBinf}) also
gives the asymptotic form for $\langle \bigl| \dot\Psi_0 \bigr|^2
\rangle^{H-U}$ and $\langle \bigl| \dot\Psi_4 \bigr|^2 \rangle^{H-U}$.

\section{NUMERICAL RESULTS}
\label{numerical}
In this section we present the results of evaluating expressions
(\ref{Sdiff}) for the
differences between renormalized expectation values of $\bigl|
\dot\Psi_A \bigr|^2$
in the Boulware, Unruh and Hartle-Hawking vacua, using values for the
coefficients $| B_{l\omega}  ^{\rm in} |^2$ and $A_{l\omega}  ^{\rm
up}$ which were obtained numerically in Sec.~\ref{ABcoefficients}.

All power series and sums over $l$ converge swiftly; likewise the
infinite
upper limit on the integral presents no difficulty, and may be
approximated
with sufficient accuracy by $(2M)\omega=2$ in every case. There are
nevertheless
technical problems associated with the computation, two of which merit
some
explanation here.

The first problem concerns the evaluation of the integrands of
(\ref{Sdiff}) at the
lower limit of the integral $\omega=0$: although the integrand of
$\langle\bigl| \dot\Psi_A \bigr|^2\rangle^{H-U}$ can be seen to vanish
at $\omega=0$ (consider
(\ref{E7})), the integrand of $\langle\bigl| \dot\Psi_A
\bigr|^2\rangle^{U-B}$ is non-zero at
$\omega=0$ and its value there can only be determined by expanding the
integrand as
a Taylor series in powers of $\omega$ \ \cite{BP}.
 To this end consider the Taylor expansions
\begin{eqnarray*}
\,{}_{-2} S_{l\omega}
&&= {}_{-2} S_{l0}
  +  [{}_{-2} S_{l0}]' \omega
  + {\cal O} \left( \omega^2 \right) \\
{}_{+2} \hat S_{l\omega}
&&\equiv \omega  \, \,{}_{+2} S_{l\omega}
={}_{+2} \hat S_{l0}
  +  [{}_{+2} \hat S_{l0}]' \omega
  + {\cal O} \left( \omega^2 \right) \\
{A_{l\omega}  ^{\rm up} \over \omega}
&&=  [  A_{l0}^{{\rm up}} ]'
  +{1 \over 2}  [ A_{l0}^{{\rm up}} ]'' \omega
  +{\cal O} \left( \omega^2 \right)
\end{eqnarray*}
where we have introduced the notation
\begin{equation}
[f_{l0}]' = \left[ {\partial f_{l\omega  } \over \partial \omega}
\right]_{\omega  =0}.
\end{equation}
We have  expanded ${}_{+2} \hat S_{l\omega}  $ rather than $\,{}_{+2}
S_{l\omega}  $
since the latter diverges like $\omega^{-1}$ as
$\omega\to 0$ (consider Eq. (\ref{recrel}) when $k=s=2$). We have also
used
 $A^{\rm up}_{l0}=0$ which follows from (\ref{E7}).
Inserting these expansions into the power series
representation (\ref{StoR:2}) for $\,{}_{+2} R_{l\omega}  ^{\rm up}$
one obtains
\begin{eqnarray}
\,{}_{+2} R_{l\omega}  ^{\rm up}=
&&\left\{ {}_{-2}S_{l0}^*
+\triangle^{-2}[A_{l0}^{{\rm up}}]' {}_{+2}\hat S_{l0} \right\}
+\Bigl\{ [{}_{-2}S_{l0}^*]'
+ir_* \, {}_{-2}S_{l0}^*   + \nonumber \\
&&+ \triangle^{-2} [A_{l0}^{{\rm up}}]'
\Bigl( [{}_{+2} \hat S_{l0}]'
-ir_* \, {}_{+2} \hat S_{l0} \Bigr) +{1 \over 2} \triangle^{-2}
 [ A_{l0}^{{\rm up}} ]''
{}_{+2} \hat S_{l0}\Bigr\} \omega
+{\cal O} \left( \omega^2 \right)  .
\label{smallomega}
\end{eqnarray}
In order that Eq. (\ref{diff:3}) for $\langle \bigl| \dot\Psi_0
\bigr|^2 \rangle^{U-B}$ be finite for
finite $r>2M$, we must have $\,{}_{+2} R_{l\omega}  ^{\rm up} \sim
\omega^n$ as $\omega\to0$ where $n\ge1$, i.e.
\begin{equation}
{}_{+2}R_{l0}^{\rm up}=0
\label{omega=0}
\end{equation}
and the value of the integrand in expression (\ref{diff:3}) for
$\langle\bigl| \dot\Psi_0 \bigr|^2\rangle^{U-B}$ at $\omega=0$ is
\begin{equation}
{1\over 2^4 \pi^2 (2M)^7} \sum_{l=2}^\infty (2l+1)
\bigl[ (l-1)l(l+1)(l+2) \bigr]^2
\Bigl|[{}_{+2}R_{l0}^{\rm up}]'\Bigr|^2
\label{omega=0sum}
\end{equation}
where $[{}_{+2}R_{l0}^{\rm up}]'$
is the coefficient multiplying $\omega$  in (\ref{smallomega}).
(\ref{omega=0sum}) can now be
evaluated numerically; in particular the power series are determined
straightforwardly from (\ref{Ss}) and (\ref{recrel}), and
$[A_{l0}^{{\rm up}}]'$ is obtained from the formula
\begin{equation}
 [A_{l0}^{{\rm up}}]'
 = {-\triangle^2 \, {}_{-2}S_{l0}^*
\over {}_{+2}\hat S_{l0}}
\label{Aupprime}
\end{equation}
which follows from (\ref{smallomega}) and (\ref{omega=0}). The second
derivative $[A_{l0}^{{\rm up}}]''=\left[ \left(\partial^2 /\partial^2
\omega
\right) \left(A_{l\omega}  ^{\rm up}\right) \right]_{\omega=0}$ is
computed using the approximation
\begin{equation}
 [A_{l0}^{{\rm up}}]''
\approx
- {2 \over {}_{+2} \hat S_{l0} }
\left[  {\partial\over\partial{\omega}}  \left(A_{l\omega}  ^{\rm
up}\right)  {\partial\over\partial{\omega}}  \left({}_{+2} \hat
S_{l\omega}
\right)    +\triangle^2 \left\{ {\partial\over\partial{\omega}}
\left(\,{}_{-2} S_{l\omega}  ^* \right)
+2ir_*\,{}_{-2}S_{l\omega}  ^* \right\}
 \right]_{\omega=0}
\label{Aupdoubleprime}
\end{equation}
(which follows from expanding both sides of (\ref{Aupapprox}) in powers
of $\omega$  and using
(\ref{Aupprime})).  Note that (\ref{Aupdoubleprime}) will only become
independent of $r$  at large
radii, in contrast with (\ref{Aupprime}) which may be computed at any
value of $r>2M$.

The value of the integrand of (\ref{diff:4}) at $\omega=0$ is found in
a similar way,
requiring only that $\left[ ({\rm d}/{\rm d}r){}_{+2}\hat S_{l\omega}
\right]_{\omega=0}$ and $\left[ ({\rm d}/{\rm d}r){}_{-2} S_{l\omega}
^*
\right]_{\omega=0}$ be evaluated numerically in addition to the power
series
considered above. In both cases the result is finite and
non-vanishing.

The second technical difficulty also arises during the computation of
$\langle\bigl| \dot\Psi_A \bigr|^2\rangle^{U-B}$, when one attempts to
evaluate $|\,{}_{s} R_{l\omega}  ^{\rm up}|^2$
numerically using the power series representations (\ref{StoR:2}),
(\ref{StoR:4}).  This latter quantity is composed of two pieces (see
e.g. the expression for $|\,{}_{+2} R_{l\omega}  ^{\rm up}|^2$ in
braces in (\ref{StoR:3})),
which turn out to be opposite in sign but equal in magnitude to high
accuracy; one must therefore work to high precision in order to produce
reliable results.  The problem becomes more severe as $\omega\to 0$.
This
difficulty can be understood by once again invoking the analogy between
our system and a classical scattering problem; at low frequencies,
upcoming radiation from the black hole is unable to surmount the
potential barrier $\,{}_{s} V_{l\omega}  (r)$ and is instead completely
reflected back across the event horizon.

\begin{figure}
\centerline{\epsfig{figure=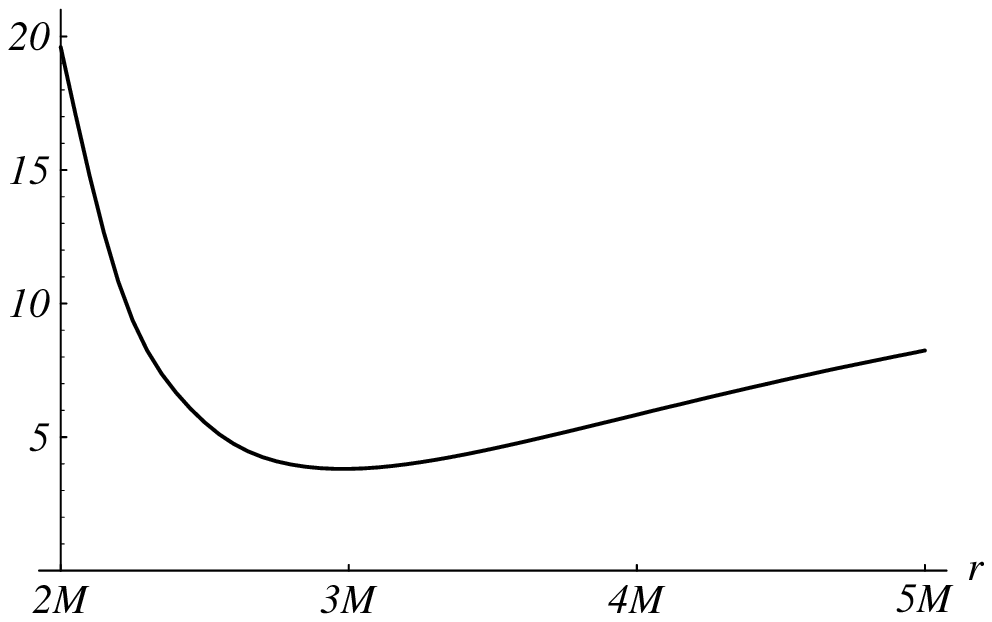,height=9.5truecm}}
\bigskip\caption{$(2M)^6 (1 - 2M/r)^4 \langle \bigl| \dot\Psi_0
\bigr|^2 \rangle^{H-U} \times  10^5$.  The value of this combination at
$r=2M$ is
$19{\cdot}6$ and its asymptotic value at infinity is $15{\cdot}8$.}
\label{pzshhmunfig}
\end{figure}

\begin{figure}
\centerline{\epsfig{figure=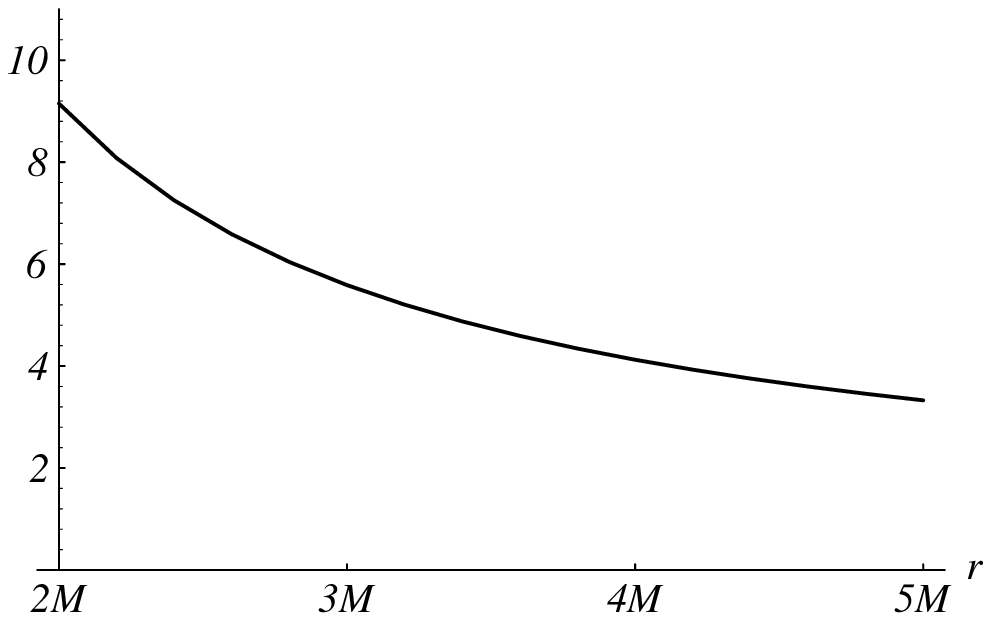,height=9.5truecm}}
\bigskip\caption{$(2M)^6 (1 - 2M/r)^{-4} \langle \bigl| \dot\Psi_4
\bigr|^2 \rangle^{H-U} \times  10^5$. The value of this combination at
$r=2M$ is
$9{\cdot}14$ and its asymptotic value at infinity is $0{\cdot}988$.}
\label{pfshhmunfig}
\end{figure}

\begin{figure}
\centerline{\epsfig{figure=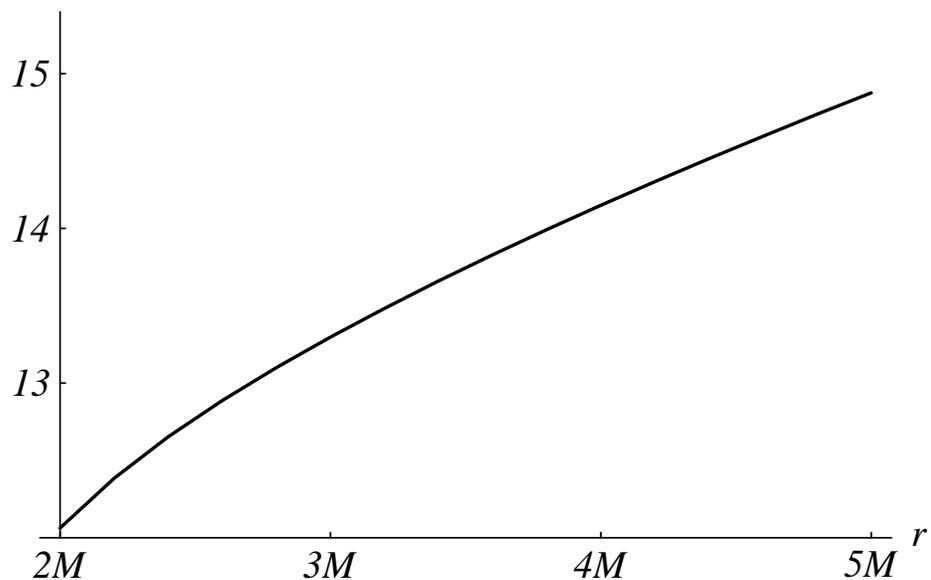,height=9.5truecm}}
\bigskip\caption{$(2M)^6 (r/2M)^{10}(1-2M/r)^5 \langle \bigl|
\dot\Psi_0 \bigr|^2
\rangle^{U-B} \times  10^3$. The value of this combination at $r=2M$ is
$12{\cdot}1$ and its asymptotic value at infinity is  $29.1$.}
\label{pzsunmbofig}
\end{figure}

\begin{figure}
\centerline{\epsfig{figure=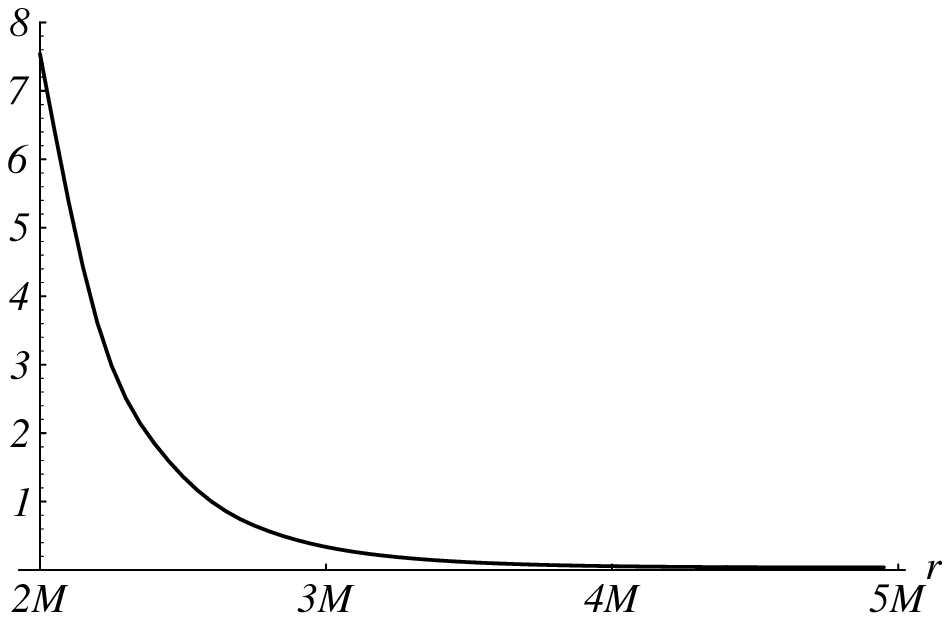,height=9.5truecm}}
\bigskip\caption{$(2M)^6 (r/2M)^2(1-2M/r) \langle \bigl| \dot\Psi_4
\bigr|^2 \rangle^{U-B} \times  10^4$.
The value of this combination at $r=2M$ is
$7{\cdot}54$ and its asymptotic value at infinity is $0{\cdot}0842$.}
\label{pfsunmbofig}
\end{figure}

\begin{figure}
\centerline{\epsfig{figure=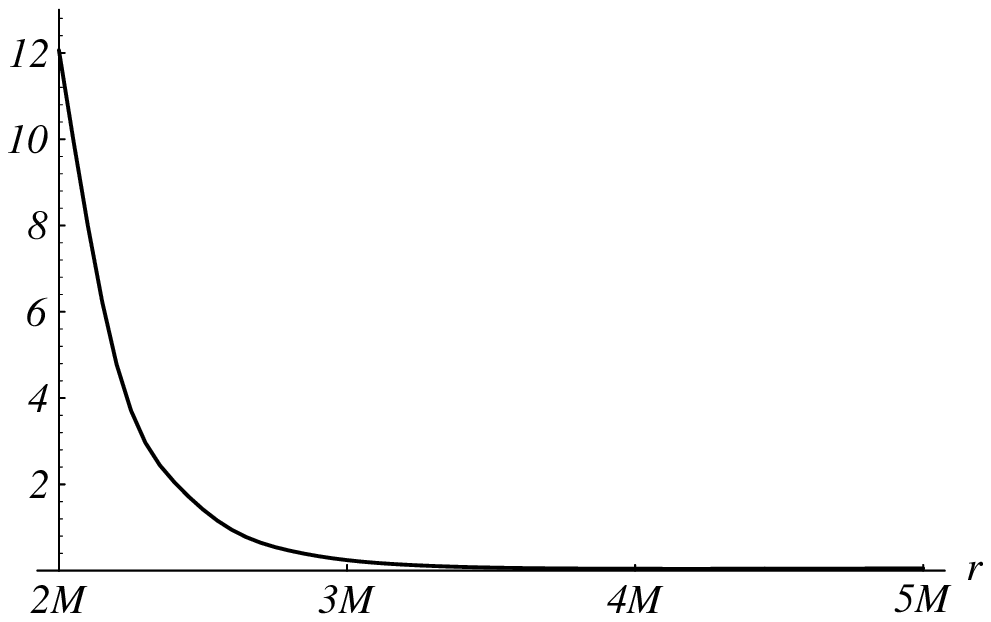,height=9.5truecm}}
\bigskip\caption{$(2M)^6 (1 - 2M/r)^5 \langle \bigl| \dot\Psi_0
\bigr|^2 \rangle^{H-B} \times  10^3 = (2M)^6 16 (1 - 2M/r) \langle
\bigl| \dot\Psi_4 \bigr|^2 \rangle^{H-B} \times  10^3$.
The value of this combination at $r=2M$ is
$12{\cdot}1$ and its asymptotic value at infinity is  $0{\cdot}158$.}
\label{pfshhmbofig}
\end{figure}

The final results are displayed graphically in Figs.
\ref{pzshhmunfig}-- \ref{pfshhmbofig}, where the quantities have been
scaled to give finite values on the horizon and at infinity.  The most
physically interesting graphs are those for $\langle\bigl| \dot\Psi_4
\bigr|^2\rangle^{U-B}$ (Fig. \ref{pfsunmbofig}) corresponding to
outgoing radiation from an evaporating black hole and $\langle\bigl|
\dot\Psi_A \bigr|^2\rangle^{H-B}$  (Fig. \ref{pfshhmbofig})
corresponding to a black hole in thermal equilibrium at the Hawking
temperature.  The most striking feature of these graphs is the very
rapid decline in vacuum activity with $r$; this is even more pronounced
when the asymptotic scaling (which softens the effect) is removed.
This suggests that
quantum gravitational effects may play a highly significant role in
determining the back-reaction near the horizon even though the
asymptotic flux to infinity measured by $\dot M$ is less that that due
to lower spin fields.

\section{CONCLUSION}
\label{conclusion}
  In Sec. \ref{powerseries} we chose to apply Leaver's method directly
  to the Teukolsky equation, this has the great virtue of directness.
  We now briefly mention an alternative procedure which we considered
  employing but rejected.   This alternative is  t

o work with the Regge-Wheeler equation
\begin{equation}
\left[{{\rm d}^2 \over {\rm d}r_*{}^2} + \omega^2 - U_{l \omega}
  \right] F_{l \omega}   = 0 ,
			       \label{RWeqn}
\end{equation}
where
\begin{equation}
 U_{l \omega}   = {r-2M \over r} \left( {l(l+1) \over r^2} - {6M \over
 r^3} \right).
\end{equation}
Eq. (\ref{RWeqn}) has independent solution $F^{\rm in}$ and $F^{\rm
up}$ with asymptotic behavior
\begin{equation}
     F^{\rm in}_{l\omega}   \sim \cases{
	     b_{l\omega}   {\rm e}^{-i\omega  r_*} ,& as $r \to 2M$,
	     \cr & \cr
	     {\rm e}^{-i\omega  r_*} + a^{\rm in}_{l \omega}
			 {\rm e}^{+i\omega  r_*} ,
					       &as $ r \to \infty$
					       ,\cr}
\end{equation}
and
\begin{equation}
     F^{\rm up}_{l\omega}   \sim \cases{
	      {\rm e}^{+i\omega  r_*} + a^{\rm up}_{l \omega}   {\rm
	      e}^{-i\omega  r_*} ,
	      & as $r \to 2M$, \cr & \cr
	  b_{l\omega}   {\rm e}^{+i\omega  r_*}   ,      &as $ r \to
	  \infty$. \cr}
\end{equation}
Defining the complex differential operator
\begin{equation}
   \overrightarrow{\cal O} \equiv 2r(r-3M+i\omega  r^2) \left( {{\rm d}
   \over {\rm d}r_*} + i \omega  \right) + r^3 U_{l \omega}   ,
\end{equation}
we can then write the Teukolsky functions in terms of  $F^{\rm in}$ and
$F^{\rm up}$ as
\begin{mathletters}
\begin{eqnarray}
   \,{}_{-2} R_{l\omega}  ^{\rm in} &=& - {4 \omega^2 \over
   C_{l\omega}  {}^*} \overrightarrow {\cal O} F^{\rm in}_{l\omega}
   \\
\,{}_{-2} R_{l\omega}  ^{\rm up}&=& {(2M)^3 \over 2 |C_{l\omega}  |^2}
(1+2iM\omega)(1+4iM\omega) \overrightarrow {\cal O}
				      F^{\rm up}_{l\omega}   \\
  \,{}_{+2} R_{l\omega}  ^{\rm in} &=& 16 \omega^2 \, \Delta^{-2}
  {\overrightarrow {\cal O}}{}^* F^{\rm in}_{l\omega}   \\
\,{}_{+2} R_{l\omega}  ^{\rm up} &=& {(2M)^3 \over C_{l\omega}  {}^*}
(1+2iM\omega)(1+4iM\omega) \, \Delta^{-2}
		     {\overrightarrow {\cal O}}{}^* F^{\rm
		     up}_{l\omega}
\end{eqnarray}
\end{mathletters}

The apparent advantage of the Regge-Wheeler approach is that it deals
with a real equation similar to the spin-0 equation.  However,
the three term recursion relation obtained through Leaver's
approach is still complex so no real advantage is accrued.  In
addition, the
expressions for the Weyl scalars involve derivatives of the
Regge-Wheeler functions which considerably complicates and obscures the
asymptotic analysis required as compared with  that employed in the
more direct
approach we have followed.

Having argued that there is no advantage in using the Regge-Wheeler
formalism for { \it our} calculation, we should point out that some of
the asymptotic formulae involving Teukolsky transmission and reflection
coefficients look far neater in terms of Regge-Wheeler transmission and
reflection coefficients.  For example, it can be shown that
\begin{equation}
    |b_{l\omega}   |^2  = {(2M)^5 p_\omega  \over 4 \omega^5} |B^{\rm
    in} _{l\omega}  |^2 =
			 {4 \omega^5 \over (2M)^5 p_\omega}   |B^{\rm
			 up} _{l\omega}  |^2,
\end{equation}
and so, Eq.~(\ref{mdot}) takes the form
\begin{equation}
\dot M_{\omega}   = 2 \times {1 \over 2 \pi} \sum_{l=2}^\infty (2l+1)
     {\omega   \left| b_{l \omega}   \right| ^2  \over
	   \left({\rm e}^{8\pi M\omega}   -1\right)}
\quad ,
\end{equation}
which parallels the scalar result except that the sum starts at $l=2$,
the lowest radiative mode of the graviton field, and there is the extra
factor of 2 arising from the number of polarisation states for the
graviton.

One might view the research presented here as a precursor to the
numerical evaluation of the renormalized effective  energy-momentum
tensor for quantized linear gravitational perturbations of a black
hole. However, there are severe problems of gauge invariance in
defining  such an object. One approach to this problem, using the
Vilkovisky-DeWitt off-shell gauge-invariant effective  action, was
suggested in Ref. \cite{AFO}.  Since both the Vilkovisky-DeWitt
effective energy-momentum tensor $\langle T_\mu{}^\nu \rangle$ and
$\langle \bigl| \dot\Psi_A \bigr|^2 \rangle$ consist of terms which are
quadratic in the
metric perturbation $h_{\mu\nu}$ and its derivatives (compare Eqs.
(\ref{psiNP}) of this paper with Eq. (2.6) of Ref. \cite{AFO}), the
evaluation of the {\it differences} in the renormalized expectation
values $\langle T_\mu{}^\nu \rangle^{H-U}$ and  $\langle T_\mu{}^\nu
\rangle^{U-B}$ will be a straightforward (though laborious) extension
of the calculation outlined in this paper. However, if one wishes to
compute an individual renormalized expectation value ($\langle H|
T_\mu{}^\nu
| H \rangle_R$ say) then new ground must be broken, since in this case
explicit renormalization is necessary. Problems then arise since the
graviton renormalization scheme of Ref. \cite{AFO} can be only
implemented  in deDonder gauge $h^\mu{}_{\nu;\mu}-{1 \over 2}
h^\mu{}_{\mu;\nu}=0$ (where the propagator has Hadamard form), whereas
a complete set of solutions to the linearized field equations currently
exists only in radiation gauge (see Eqs.  (\ref{h})).  If these
technical problems can be resolved, it would be interesting to compare
this quantity with its scalar and electromagnetic analogues.   To date
it has only been possible to make quantitative comparisions of the
black hole luminosity of gravitons with that due to radiation of
massless particles of lower  spin \cite{P}.  One finds that the
luminosity for gravitons is less than for lower spin fields but one
expects that this is due to the increase in the height of the effective
potential barrier with spin and so  cannot be used to draw conclusions
about their importance near the black hole.  Indeed, we expect that as
in the case of quantum fields propagating near a conical
singularity,\cite{AMO} the contribution of gravitons will be seen to
dominate the back-reaction near a spherical singularity (see also the
remarks at the end of Sec. \ref{numerical}).

 In this paper we have chosen to sidestep these gauge-invariance
problems and instead, following Ref. \cite{CCH}, have concentrated on
the gauge-independent Newman-Penrose scalars $\dot\Psi_0$ and
$\dot\Psi_4$.
These scalars are in some sense the true gravitational field variables
in the Schwarzschild background, for example, it is in principle
possible to reformulate the quantum theory at the level of the one-loop
effective action in terms of $\dot\Psi_0$ and $\dot\Psi_4$ (strictly
speaking,
$\dot\Psi_0$ may be regarded as representating the two radiative
degrees of
freedom and as acting as a `superpotential' from which $\dot\Psi_4$ may
be
obtained). Most importantly, the expectation values we have computed
provide important physical measures of the vacuum activity of the
quantized gravitational field around a black hole and, in the
asymptotic regimes, they directly measure the one-loop quantum
gravitational
energy flux across the horizon of the black hole and to infinity.

\appendix
\section{NOTATION AND NP CONVENTIONS}
\label{notation}

 For convenience we repeat  here the definition of various symbols used
 in the text:
\begin{eqnarray*}
 \triangle&=& r(r-2M)\\
 C_{l\omega} &=& (l-1)l(l+1)(l+2) + 12iM\omega  \\
  p_\omega  &=& 2M
\omega  ( 1 +4 M^2 \omega^2 ) ( 1 + 16 M^2 \omega^2) .
 \end{eqnarray*}

The following are the conventions and notation employed in this paper
for the
NP description of general relativity.
In the NP formalism the geometry of a general space-time
with metric $g_{\mu\nu}(x)$ is encoded into a null complex tetrad
$\left\{e_{(a)}{}^\mu(x):a=1,2,3,4\right\}$ which satisfies the
orthonormality
conditions
\[
g_{\mu\nu} e_{(a)}{}^\mu e_{(b)}{}^\nu = \eta_{(a)(b)} \quad .
\]
$\eta_{(a)(b)}$ is the constant symmetric matrix
\[
\eta_{(a)(b)}=\left[\matrix{\phantom{-}0&-1&0&0\cr
			  -1&\phantom{-}0&0&0\cr
			  \phantom{-}0&\phantom{-}0&0&1\cr
			  \phantom{-}0&\phantom{-}0&1&0\cr}\right]_{(a)(b)}
\]
and acts as the NP analogue of the metric tensor; in particular tetrad
indices
are lowered and raised by $\eta_{(a)(b)}$ and its matrix inverse,
denoted
$\eta^{(a)(b)}$, respectively. We adopt the convention whereby tensor
indices are
labelled by Greek letters and tetrad indices by Roman letters enclosed
in
parentheses.

By expressing tensors
$T_{\ldots\mu\ldots}$ in terms of their {\it tetrad
components} $T_{\ldots (a)\ldots}=e_{(a)}{}^\mu
T_{\ldots\mu\ldots}$ and introducing the {\it spin
coefficients} (NP analogues of Christoffel symbols)
\[
\gamma_{(a)(b)(c)}=e_{(a)}{}^\beta{}_{;\gamma}\, e_{(b)\beta}\,
e_{(c)}{}^\gamma\quad,
\]
the fundamental equations of general relativity (e.g.~Einstein's
equation, the
Bianchi identities) can all be expressed as tetrad equations involving
only
scalar quantities.

We also adopt the notation (see e.g. Ref.~\cite{Ch})
\[
e_{(a)}{}^\mu \partial_\mu \equiv (D,\Delta,\delta,\delta^*)
\]
for the tetrad operators,
\begin{eqnarray*}
&&\kappa\equiv\gamma_{(3)(1)(1)}\qquad
\rho\equiv\gamma_{(3)(1)(4)}\qquad
\epsilon\equiv{1 \over
2}\left(\gamma_{(2)(1)(1)}+\gamma_{(3)(4)(1)}\right)\\
&&\sigma\equiv\gamma_{(3)(1)(3)}\qquad
\mu\equiv\gamma_{(2)(4)(3)}\qquad
\gamma\equiv{1 \over
2}\left(\gamma_{(2)(1)(2)}+\gamma_{(3)(4)(2)}\right)\\
&&\lambda\equiv\gamma_{(2)(4)(4)}\qquad
\tau\equiv\gamma_{(3)(1)(2)}\qquad
\alpha\equiv{1 \over
2}\left(\gamma_{(2)(1)(4)}+\gamma_{(3)(4)(4)}\right)\\
&&\nu\equiv\gamma_{(2)(4)(2)}\qquad
\pi\equiv\gamma_{(2)(4)(1)}\qquad
\beta\equiv{1 \over
2}\left(\gamma_{(2)(1)(3)}+\gamma_{(3)(4)(3)}\right)
\end{eqnarray*}
for the spin coefficients, and
\begin{eqnarray*}
&&\Psi_0\equiv-C_{(1)(3)(1)(3)}\qquad
\Psi_1\equiv-C_{(1)(2)(1)(3)}\qquad
\Psi_2\equiv-C_{(1)(3)(4)(2)}\\
&&\hfil\Psi_3\equiv-C_{(1)(2)(4)(2)}\qquad
\Psi_4\equiv-C_{(2)(4)(2)(4)}\hfil
\end{eqnarray*}
for the five independent tetrad components of the Weyl tensor,   called
the {\it Weyl scalars}.

\section{ SPIN-WEIGHTED SPHERICAL HARMONICS}
\label{harmonics}
{\it Spin-weighted spherical harmonics} $\,{}_{s} Y_l^m(\theta,\phi)$
are defined for
$s=-l,-l+1,\ldots,l-1,l$ by the set of equations (Refs~\cite{NP},
\cite{G})
\begin{mathletters}
\label{B1}
 \begin{eqnarray}
\,{}_{0} Y_l^m(\theta,\phi)&&=Y_l^m(\theta,\phi)  \label{B1:1} \\
\,{}_{s+1} Y_l^m(\theta,\phi)&&=\left[(l-s)(l+s+1)\right]^{-{1 \over
2}}\hbox{$\partial$\kern -4pt ${}^{-}$}\,{}_{s} Y_l^m(\theta,\phi)
\label{B1:2}\\
\,{}_{s-1} Y_l^m(\theta,\phi)&&=-\left[(l+s)(l-s+1)\right]^{-{1 \over
2}}\hbox{$\partial$\kern -4pt ${}^{-}$}'\,{}_{s} Y_l^m(\theta,\phi)
\label{B1:3}
\end{eqnarray}
\end{mathletters}
in terms of the ordinary spherical
harmonics
\[
Y_l^m(\theta,\phi) =
\left[{2l+1\over4\pi}{(l-|m|)!\over(l+|m|)!}\right]^{1 \over 2}
P_l^{|m|}\left(\cos\theta\right) {\rm e}^{im\phi} \quad .
\]
Here
$\hbox{$\partial$\kern -4pt ${}^{-}$}$
and $\hbox{$\partial$\kern -4pt ${}^{-}$}'$ are operators which act as
follows on a quantity $\eta$ of
spin-weight $s$:
\begin{mathletters}
\label{B2}
\begin{eqnarray}
\hbox{$\partial$\kern -4pt ${}^{-}$}\eta
&&=-\sin^s\theta \left[{\partial\over\partial{\theta}}
+{i\over\sin\theta}{\partial\over\partial{\phi}} \right]
\left(\left( \sin^{-s}\theta \right) \eta \right) \label{B2:1} \\
\hbox{$\partial$\kern -4pt ${}^{-}$}'\eta
&&=-\sin^{-s}\theta \left[ {\partial\over\partial{\theta}}
-{i\over\sin\theta}{\partial\over\partial{\phi}}  \right]
\left(\left( \sin^{s}\theta \right) \eta \right).
\label{B2:2}
\end{eqnarray}
\end{mathletters}
 The application of $\hbox{$\partial$\kern -4pt ${}^{-}$}$ on a
 quantity  lowers the spin-weight by 1:
the application of $\hbox{$\partial$\kern -4pt ${}^{-}$}'$ on a
quantity raises   the spin-weight by 1.
It follows immediately from (\ref{B1}) that
\begin{equation}
\hbox{$\partial$\kern -4pt ${}^{-}$}'\hbox{$\partial$\kern -4pt
${}^{-}$}\,{}_{s} Y_l^m(\theta,\phi)=-(l-s)(l+s+1)\,{}_{s}
Y_l^m(\theta,\phi)
\end{equation}
or, substituting (\ref{B2}) for $\hbox{$\partial$\kern -4pt ${}^{-}$}$
and $\hbox{$\partial$\kern -4pt ${}^{-}$}'$, that
\begin{eqnarray*}
\left[
{1\over\sin\theta} {\partial\over\partial{\theta}}  \left( \sin\theta
{\partial\over\partial{\theta}}  \right)
+{1\over\sin^2\theta} {\partial^2\over\partial^2{\phi}} \right.
&&+{2is\cos\theta\over\sin^2\theta} {\partial\over\partial{\phi}}    \\
&& \left. -s^2\cot^2\theta + s + (l-s)(l+s+1) \right]
\,{}_{s} Y_l^m(\theta,\phi) = 0.
\end{eqnarray*}
The spin-weighted spherical harmonics satisfy the orthonormality and
completeness relations (Refs.\cite{NP}, \cite{G}):
\begin{equation}
\int_0^{2\pi} {\rm d}\phi \int_0^\pi \sin\theta \, {\rm d}\theta
\,{}_{s} Y_l^m(\theta,\phi) \, {}_s Y_{l'}^{m'*}(\theta,\phi)
= \delta_{ll'} \delta_{mm'}
\label{B3}
\end{equation}
and
\begin{equation}
\sum_{l=s}^\infty \> \sum_{m=-l}^l \,{}_{s} Y_l^m(\theta,\phi) \,{}_s
Y_l^{m*}(\theta',\phi')
=\delta(\phi-\phi') \delta(\cos\theta-\cos\theta')
\end{equation}
respectively; these follow essentially by induction on
 (\ref{B1}) and the corresponding relations for the
ordinary spherical harmonics. Using the same method one can verify that
the
``addition theorem''
\begin{equation}
\sum_{m=-l}^l \left| \,{}_{s} Y_l^m(\theta,\phi) \right|^2
= {2l+1\over4\pi}
\label{B4}
\end{equation}
holds for $s=\pm1,\pm2$ as well as $s=0$.\cite{V}

\section{ ASYMPTOTIC EXPANSIONS OF THE RADIAL FUNCTIONS AT INFINITY}
\label{asymptoticapp}
Based on (\ref{rlims:1}), $\,{}_{-2} R_{l\omega}  ^{\rm in}$ will have
an asymptotic expansion of the form
\begin{equation}
\,{}_{-2} R_{l\omega}  ^{\rm in} \sim
{\rm e}^{-i\omega  r_*} \left\{ {1\over r} +{\cal O} \left( {1\over
r^2} \right) \right\}
+ A_{l\omega}  ^{\rm in} {\rm e}^{+i \omega  r_*} \left\{ r^3 +b_1 r^2
+b_2 r +b_3 +{b_4\over r}
+{\cal O}\left({1\over r^2}\right) \right\} \quad.
\label{C1}
\end{equation}
The coefficients $b_n(l,\omega)$ are determined by substituting this
expansion into the $s=-2$ radial equation  (\ref{req}) and equating the
coefficient of each power of $r$ to zero; we find
\begin{eqnarray*}
b_1={(l-1)(l+2)i\over 2 \omega}  \quad\quad
&&b_2=- {C_{l\omega}^*\over 8\omega^2}           \\
b_3=- {l(l+1)iC_{l\omega}^*\over 48\omega^3}\quad\quad
&&b_4={|C_{l\omega}|^2\over 384\omega^4} \quad.
\end{eqnarray*}
Next by inserting (\ref{C1}) into (\ref{+2to-2}) we deduce that
\begin{equation}
\,{}_{+2} R_{l\omega}  ^{\rm in} \sim
64 \omega^4 {{\rm e}^{-i\omega  r_*} \over r}
+{ |C_{l\omega}|^2 \over 4 \omega^4} A_{l\omega}  ^{\rm in}
{{\rm e}^{+i \omega  r_*}\over r^5}
\label{C2}
\end{equation}
in the limit as $r\to\infty$. A similar argument yields the following
asymptotic expansions for the $s=+2$ radial functions:
\begin{eqnarray}
\,{}_{+2} R_{l\omega}  ^{\rm up}
&&\sim
B_{l\omega}  ^{\rm up} {\rm e}^{+i \omega  r_*} \left\{ {1\over r^5}
+{c_1\over r^6} +{c_2\over r^7}
+{c_3\over r^8} +{c_4\over r^9} +{\cal O}\left({1\over r^{10}}\right)
\right\}
\label{C3}   \\
\,{}_{-2} R_{l\omega}  ^{\rm up}
&&\sim
B_{l\omega}  ^{\rm up} {\rm e}^{+i \omega  r_*} \left\{ d_1 r^3 +d_2
r^2 +d_3 r +d_4
+{\cal O} \left({1\over r}\right) \right\}
\label{C4}
\end{eqnarray}
as $r \to \infty$, where
\begin{eqnarray*}
c_1
&&= {i\over 2\omega}   \left[ (l-2)(l+3) \right] \\
c_2
&&= {i\over 4\omega}   \left[ \left\{(l-3)(l+4) - 4M\omega  i
\right\}c_1
+2M \left\{ 15 - (l-2)(l+3) \right\} \right]  \\
c_3
&&= {i\over 6\omega}   \left[ \left\{(l-4)(l+5) - 8M\omega  i
\right\}c_2
+2M \left\{ 30-(l-2)(l+3) \right\}c_1
-15(2M)^2 \right]  \\
c_4
&&= {i\over 8\omega}   \left[ \left\{(l-5)(l+6) -12M\omega  i
\right\}c_3
+2M\left\{49-(l-2)(l+3) \right\}c_2
-24(2M)^2 c_1 \right]
\end{eqnarray*}
and
\begin{eqnarray*}
d_1 = {4 \omega^4\over  \left|C_{l\omega}\right|^2 } \quad\quad
&&d_2 = {2(l-1)(l+2)i\omega^3 \over \left|C_{l\omega}\right|^2 }  \\
d_3 = - {\omega^2 \over 2C_{l\omega}} \quad\quad\quad
&&d_4 =  - { l(l+1)i\omega   \over 12C_{l\omega}}
\end{eqnarray*}
Note that (\ref{C2}), (\ref{C4}) are consistent with
(\ref{simpleasymp}).

\section{RELATIONS BETWEEN THE REFLECTION AND
TRANSMISSION COEFFICIENTS}
\label{relations}
 Eq~(\ref{scateq}) gives the spin $s$ radial equation
 in Liouville normal form. The Wronskian
\[
W \left[ \,{}_{s} Q_{l\omega}  (r),{}_s\tilde{Q}_{l\omega}  (r) \right]
\equiv
\,{}_{s} Q_{l\omega}  (r) {{\rm d}\over {\rm d}{r_*}}
{}_s\tilde{Q}_{l\omega}  (r)
-{}_s\tilde{Q}_{l\omega}  (r) {{\rm d}\over {\rm d}{r_*}}  \,{}_{s}
Q_{l\omega}  (r)
\]
of any two particular
solutions $\,{}_{s} Q_{l\omega}  (r),{}_s\tilde{Q}_{l\omega}  (r)$ of
the equation will be constant.
In addition, observe  from  (\ref{pot}) that
$\,{}_{-s} V_{l\omega}  ^*(r)=\,{}_{s} V_{l\omega}  (r)$; thus if
$\,{}_{s} Q_{l\omega}  (r)$ is a solution of (\ref{scateq}), so is
$\,{}_{-s} Q_{l\omega}  ^*(r)$. Hence
\[
W \left[ \,{}_{s} Q_{l\omega}  (r), \,{}_{-s} Q_{l\omega}  ^*(r)\right]
\]
will also be constant.
These facts can be used to derive
relations between the amplitudes $A_{l\omega}  ^{\Lambda}$ and
$B_{l\omega}  ^{\Lambda}$.

Consider for example
\[
W\left[\,{}_{-2} Q_{l\omega}  ^{\rm in}(r),\,{}_{-2} Q_{l\omega}
^{{\rm up}}(r)\right]
\]
where we are using the obvious notation
\begin{equation}
\,{}_{s} Q_{l\omega}  ^\Lambda(r) \equiv
r\triangle^{s\over2}\,{}_{s} R_{l\omega}  ^\Lambda(r).
\end{equation}
This Wronskian is first evaluated on the horizon, by substituting the
power
series expansions (4.7a,d) for $\,{}_{-2} R_{l\omega}  ^{\rm in}$ and
$\,{}_{-2} R_{l\omega}  ^{\rm up}$ respectively,
computing the derivatives, and then taking the $r\to2M$ limit: we find
\begin{equation}
\lim_{r\to2M} \, W\left[\,{}_{-2} Q_{l\omega}  ^{\rm in}(r),\,{}_{-2}
Q_{l\omega}  ^{{\rm up}}(r)\right]
=
{ 2i (2M)^5 p_\omega  B_{l\omega}  ^{\rm in} \over
\left|C_{l\omega}\right|^2 }
\end{equation}
where $p_\omega$  is given by (\ref{pomega}).
Next we evaluate the same Wronskian in the limit as $r\to\infty$, using
the
asymptotic expansions (\ref{C1}), (\ref{C4}) derived in Appendix
\ref{asymptoticapp}
for $\,{}_{-2} R_{l\omega}  ^{\rm in}$, $\,{}_{-2} R_{l\omega}  ^{\rm
up}$ respectively. The result is
\begin{equation}
\lim_{r\to\infty} \, W\left[\,{}_{-2} Q_{l\omega}  ^{\rm
in}(r),\,{}_{-2} Q_{l\omega}  ^{{\rm up}}(r)\right]
=
 { 8i \omega^5 B_{l\omega}  ^{\rm up} \over  \left|C_{l\omega}\right|^2
 } \quad .
\end{equation}
Constancy of the Wronskian therefore yields the following relation
between the
`up' and `in' transmission amplitudes:
\begin{equation}
B_{l\omega}  ^{\rm up} = {(2M)^5 p_\omega  \over 4 \omega^5}
B_{l\omega}  ^{\rm in} \quad .
\label{D1}
\end{equation}

A similar analysis of the Wronskians
$W\left[\,{}_{-2} Q_{l\omega}  ^{\rm in}(r),\,{}_{+2} Q_{l\omega}
^{{\rm in}*}(r)\right]$ and
$W\left[\,{}_{-2} Q_{l\omega}  ^{\rm in}(r),\,{}_{+2} Q_{l\omega}
^{{\rm up}*}(r)\right]$ serves to complete the set of
independent relations between the reflection and transmission
coefficients,
yielding
\begin{equation}
{\left|C_{l\omega}\right|^2 \over (2\omega)^8}
\left|A_{l\omega}  ^{\rm in}\right|^2
+{(2M)^5 p_\omega  \over 4\omega^5}
\left|B_{l\omega}  ^{\rm in}\right|^2 = 1
\label{D2}
\end{equation}
and
\begin{equation}
(-2i\omega) A_{l\omega}  ^{\rm in} B_{l\omega}  ^{{\rm up}*}
+4M\left( 1- 2iM\omega  \right) B_{l\omega}  ^{\rm in} A_{l\omega}
^{{\rm up}*} =0
\label{D3}
\end{equation}
respectively.
All other relations between reflection and transmission coefficients
follow
from the above three.

Eq.~(\ref{D2}) may also be derived from conservation of
energy flux,
\begin{equation}
 {{\rm d} E^{\rm in}_\omega  \over {\rm d}t}
-{{\rm d} E^{\rm out}_\omega  \over {\rm d}t}
={{\rm d} E^{\rm hor}_\omega  \over {\rm d}t} \quad.
\end{equation}
To see this, first substitute (\ref{psieqs}) for $\dot \Psi_{A}^{\rm
in} (l,m,\omega, P;x)$ in  (\ref{ehor}),
 (\ref{e}) and integrate over the solid angle; next replace
 $\,{}_{\pm2} R_{l\omega}  ^{\rm in}$ by formulae
describing their leading behavior in the limits $r\to2M$, $r\to\infty$
as
appropriate; finally substitute into the conservation
equation above to obtain (\ref{D2}).

\section{SMALL $\omega$  APPROXIMATIONS FOR $A_{\lowercase{l}\omega}
^{\rm \lowercase{up}}$, $B_{\lowercase{l}\omega}  ^{\rm
\lowercase{in}}$}
\label{smallomegaapp}
In this Appendix expressions are derived for the leading behavior of
$A_{l\omega}  ^{\rm up}$ and $B_{l\omega}  ^{\rm in}$ as $\omega$
tends to zero. Our method  is analogous to
that employed by Page\cite{P} for his investigation of particle
emission rates
from a Kerr black hole, and in particular makes use of approximate
hypergeometric solutions of the spin $s$ radial equation obtained by
Churilov
and Starobinski\u\i$\;$.\cite{CS}

In terms of the dimensionless quantities $x=(r/2M-1)$ and $k=2M\omega$
, the spin
$s$ radial equation (\ref{req}) may be approximated when $k\ll1$ by
\begin{eqnarray}
&&\left[ x^2 (x+1)^2 {{\rm d}^2\over {\rm d}{x}^2}
+(s+1) x(x+1)(2x+1) {{\rm d}\over {\rm d}{x}}
+k^2 x^4 \right.\nonumber \\
&&\left. +2isk x^3
-(l-s)(l+s+1) x(x+1)
-isk(2x+1) +k^2 \right] \,{}_{s} R_{l\omega}   = 0 .
\label{E1}
\end{eqnarray}

In the region $x\ll(l+1)/k$ (which includes the horizon) the third
and fourth terms can be neglected, so that the equation has three
regular
singular points and its general solution  is expressible in terms of
hypergeometric functions:
\begin{eqnarray}
\,{}_{s} R_{l\omega}
&&= C_1\> {}_2F_1 \left(-l-s,l-s+1;1-s-2ik;-x \right) x^{-s-ik}
(x+1)^{-s+ik} \nonumber \\
&&+ C_2\> {}_2F_1 \left(-l+2ik,l+1+2ik;1+s+2ik;-x \right)
(-1)^s x^{ik} (x+1)^{-s+ik}
\label{E2}
\end{eqnarray}
where $C_1$ and $C_2$ are
constants. In the region $x\gg k+1$ (which stretches to
infinity) the last two terms of (\ref{E1}) are ignorable and  the
following general solution of the approximate equation can be deduced:
\begin{eqnarray}
\,{}_{s} R_{l\omega}
&&= D_1\> {}_1F_1 \left(l+1-s,2l+2;2ikx\right) {\rm e}^{-ikx} x^{l-s}
\nonumber \\
&&+ D_2\> {}_1F_1\left(-l-s,-2l;2ikx\right)
{\rm e}^{-ikx} x^{-l-s-1} \quad .
\label{E3}
\end{eqnarray}
One can now match these general solutions in the region of overlap
 $k+1 \ll x\ll(l+1)/k$.
(The hypergeometric functions in (\ref{E2}) may be approximated
when $x\gg k+1$  using Eq. 2 on p.108 of Ref.~\cite{EMOT}, and when
$x\ll(l+1)/k$ those in (\ref{E3}) can be simply replaced
by 1.) The following matching relations between
$C_1, C_2$ and $D_1, D_2$ obtain:
\begin{mathletters}
\label{E4}
\begin{eqnarray}
D_1
&&={\Gamma (2l+1) \over \Gamma (l-s+1)}
{\Gamma (1-s-2ik) \over \Gamma (1+l-2ik)} C_1
+ (-1)^s
{\Gamma (2l+1) \over \Gamma (l+s+1)}
{\Gamma (1+s+2ik) \over \Gamma (1+l+2ik)} C_2 \label{E4:1}\\
D_2
&&={\Gamma (-2l-1) \over \Gamma (-l-s)}
{\Gamma (1-s-2ik) \over \Gamma (-l-2ik)} C_1
+ (-1)^s
{\Gamma (-2l-1) \over \Gamma (s-l)}
{\Gamma (1+s+2ik) \over \Gamma (-l+2ik)} C_2 \quad .
\label{E4:2}
\end{eqnarray}
\end{mathletters}

Consider now the particular solution $\,{}_{-2} R_{l\omega}  ^{\rm
in}(r)$ of the $s=-2$ radial
equation. First comparing the form of this solution as $r\to2M$ (see
(\ref{rlims:1})) with
that of the general solution (\ref{E2}) in the same limit, one observes
that
\begin{equation}
C_1 = (2M)^4 B_{l\omega}  ^{\rm in} \qquad \hbox{and} \qquad C_2 = 0
\label{E5}
\end{equation}
for this particular choice. Secondly comparing the form of $\,{}_{-2}
R_{l\omega}  ^{\rm in}(r)$
 with that of (\ref{E3}) as $r\to\infty$ (see (\ref{rlims:1}) and Eq.
 2, p.278
of Ref.~\cite{EMOT} for the asymptotic expansion of the confluent
hypergeometric function ${}_1F_1$) yields
\begin{mathletters}
\label{E6}
\begin{eqnarray}
{\Gamma (2l+2) \over \Gamma (l-1)} (-2ik)^{-l-3} D_1
+ {\Gamma (-2l) \over \Gamma (-l-2)} (-2ik)^{l-2} D_2
&&= {1 \over 2M}  \label{E6:1}\\
{\Gamma (2l+2) \over\Gamma (l+3)} (2ik)^{-l+1} D_1
+{\Gamma (-2l) \over \Gamma (-l+2)} (2ik)^{l+2} D_2
&&= (2M)^3 A_{l\omega}  ^{\rm in} \quad.
\label{E6:2}
\end{eqnarray}
\end{mathletters}
Eliminating $C_1, C_2, D_1$ and $D_2$ between (\ref{E4}) (with $s=-2$),
(\ref{E5}) and
(\ref{E6}) results in a pair of  equations which determine the
incoming reflection and transmission amplitudes; the leading behavior
of
$B_{l\omega}  ^{\rm in}$ as $2M\omega\to0$ is then found to be
\begin{equation}
B_{l\omega}  ^{\rm in} \approx (2M)^{-5} {l! \over 2!} {(l-2)! (l+2)!
\over
(2l+1)! (2l)!} (-4iM\omega)^{l+3}
\quad .
\label{E7}
\end{equation}

A similar comparison of $\,{}_{+2} R_{l\omega}  ^{\rm up}$ with the
approximate hypergeometric solutions yields
\begin{equation}
A_{l\omega}  ^{\rm up} \approx 2 (2M)^4 {(l-2)! \over (l+2)!}
(-4iM\omega)
\label{E8}
\end{equation}
for small $2M\omega$ .

\begin{figure}
\centerline{\epsfig{figure=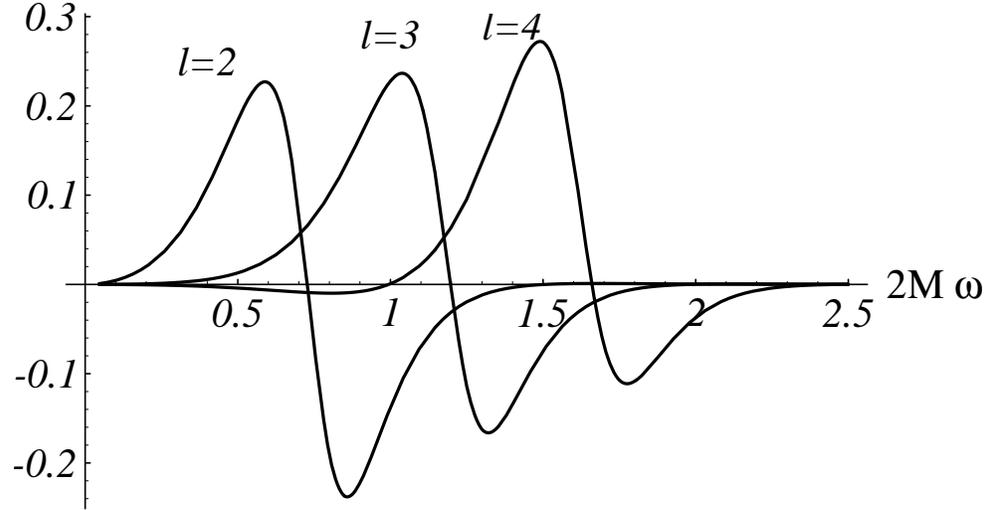,height=10truecm}}
\bigskip\caption{ The real part of $(2M)^{-4} A_{l\omega}^{\rm up}$  as
a function of
$2M\omega$  for $l=2,3,4$. }
\end{figure}

\begin{references}
\bibitem{BP}
The fact that some of the functions involved in these expansions have a
branch point at $\omega =0$ throws doubt on the validity of this naive
analysis.  However, a more careful analysis based on Appendix A of
\cite{Ca}
shows that the result is justified.
\end{references}
\end{document}